# SEMICIRCLE LAW ON SHORT SCALES AND DELOCALIZATION OF EIGENVECTORS FOR WIGNER RANDOM MATRICES

By László Erdős, Benjamin Schlein[1] and Horng-Tzer Yau[2]

*University of Munich, University of Munich and Harvard University*

We consider $N \times N$ Hermitian random matrices with i.i.d. entries. The matrix is normalized so that the average spacing between consecutive eigenvalues is of order $1/N$. We study the connection between eigenvalue statistics on microscopic energy scales $\eta \ll 1$ and (de)localization properties of the eigenvectors. Under suitable assumptions on the distribution of the single matrix elements, we first give an upper bound on the density of states on short energy scales of order $\eta \sim \log N/N$. We then prove that the density of states concentrates around the Wigner semicircle law on energy scales $\eta \gg N^{-2/3}$. We show that most eigenvectors are fully delocalized in the sense that their $\ell^p$-norms are comparable with $N^{1/p-1/2}$ for $p \geq 2$, and we obtain the weaker bound $N^{2/3(1/p-1/2)}$ for all eigenvectors whose eigenvalues are separated away from the spectral edges. We also prove that, with a probability very close to one, no eigenvector can be localized. Finally, we give an optimal bound on the second moment of the Green function.

**1. Introduction.** Denote the $(ij)$th entry of an $N \times N$ matrix $H$ by $h_{ij}$. We shall assume that the matrix is Hermitian, that is, $h_{ij} = \overline{h_{ji}}$. These matrices form a *Hermitian Wigner ensemble* if

$$(1.1) \quad h_{ij} = N^{-1/2}[x_{ij} + \sqrt{-1}y_{ij}] \qquad (i < j) \quad \text{and} \quad h_{ii} = N^{-1/2}x_{ii},$$

where $x_{ij}, y_{ij}$ $(i < j)$ and $x_{ii}$ are independent real random variables with mean zero. We assume that $x_{ij}, y_{ij}$ $(i < j)$ all have a common distribution $\nu$ with variance $1/2$ and with a strictly positive density function: $d\nu(x) = (\text{const})e^{-g(x)}\,dx$. The diagonal elements, $x_{ii}$, also have a common distribution, $d\tilde{\nu}(x) = (\text{const})e^{-\tilde{g}(x)}\,dx$, that may be different from $d\nu$. We remark

---

Received December 2007.

[1]Supported in part by the Sofja-Kovalevskaya Award of the Humboldt Foundation.

[2]Supported in part by NSF Grant DMS-06-02038.

that the special case $g(x) = x^2$ and $\tilde{g}(x) = x^2/2$ is called the *Gaussian Unitary Ensemble* (*GUE*). Let $\mathbb{P}$ and $\mathbb{E}$ denote the probability and the expectation value, respectively, w.r.t. the joint distribution of all matrix elements.

E. Wigner has introduced random matrices to model Hamiltonians, $H$, of atomic nuclei. Lacking precise knowledge about the interaction among different quantum states, he assumed that the matrix elements $(\varphi, H\psi)$ for any two orthogonal states $\varphi, \psi$ are identically distributed and as maximally independent as the unitary symmetry group acting on the Hilbert space of states allows. These assumptions already imply that the distribution of $H$ is GUE (modulo changing the expectation value and the variance). Astonishingly, this simple model very accurately reproduced the energy level statistics of various large nuclei.

Random Hamiltonians are also used in solid state physics to study electrons in disordered metallic lattices. The simplest example is the *Anderson model* on a discrete lattice, where the disorder is modeled by i.i.d. on-site potentials. The Anderson model can be generalized to continuous space and to include magnetic fields. These models are commonly referred to as *random Schrödinger operators*. Their key feature is that they have an underlying spatial structure and only matrix elements connecting nearby sites are nonzero, in contrast to the mean-field character of the Wigner ensembles.

The conductance properties of metallic lattices are strongly influenced by the spatial localization of the eigenfunctions of the corresponding Hamiltonian. Depending on the energy range, on the disorder strength and on the spatial dimension, random Schrödinger operators are believed to exhibit a transition between localized ($L^2$-normalizable) and delocalized eigenstates. These two regimes can also be characterized by the pure point or absolutely continuous spectrum, respectively. While the localization regime is fairly well understood, it remains an outstanding open problem to prove the existence of the delocalization regime. An even more ambitious conjecture states that the level spacing statistics of consecutive eigenvalues (of the finite dimensional approximation) of the random Schrödinger operator also characterizes these two regimes. In the localization regime, consecutive eigenvalues should be independent and should follow the statistics of a Poisson point process. In the delocalization regime, the level spacing statistics is believed to be identical to that of the GUE.

Random matrices are mostly studied from the point of view of eigenvalue statistics such as density of states (e.g., Wigner semicircle law) or level statistics of consecutive eigenvalues. The density of states is well understood for general Wigner matrices on macroscopic energy windows where the number of eigenstates is proportional to $N$. In our normalization this corresponds to energy windows of order one. On the finest energy scale of order $1/N$, where individual eigenvalues are observed, a universal level spacing distribution is believed to emerge that is called the Wigner–Dyson statistics. This has been proven only for Gaussian and related models (see [6] and references therein) and for Wigner matrices where the distribution of the matrix elements were Gaussian convolutions [10]. The proofs use explicit formulae for the eigenvalue correlation functions which are available only for Gaussian related models.

The eigenvalue distribution of general Wigner matrices is poorly understood on microscopic energy scales $\eta \ll 1$ due to the lack of explicit formulae for the eigenvalue distribution. The fluctuations of the density of states are known to be negligible down to energy windows of order $N^{-1/2}$ [[8], [9]] and the expected value is also known to follow the semicircle law on scales $N^{-1/2}$ and larger [2]. Under somewhat different moment assumptions, the semicircle law was shown on scales $\gg N^{-1/4}$ in [11] and the fluctuation around its mean was proven to be Gaussian in [4]. It is an open problem to show that both the fluctuation and the expected value of the density of states can be controlled down to energy scales of order $1/N$. This would be the first step toward the proof of Wigner–Dyson universality for Wigner matrices. Moreover, given the presumed connection between eigenvalue statistics and eigenfunction localization in the case of random Schrödinger operators, it is natural to investigate the (de)localization properties of the eigenvectors of random matrices. Due to the mean field character of the Wigner matrix, the eigenvectors are believed to be extended, a conjecture that is consistent with the expected repulsion of neighboring eigenvalues.

We remark that in finite dimensional Hilbert spaces extended states are characterized by $\ell^p$-norms with $p \neq 2$ instead of the absolute continuity of the spectrum. If all components of an $\ell^2$-normalized vector $\mathbf{v} \in \mathbb{C}^N$ are equal, then $\|\mathbf{v}\|_p = N^{1/p-1/2}$. Thus, deviations of the $\ell^p$-norm of an eigenvector from $N^{1/p-1/2}$ can be used to quantify the delocalization properties of the state. In particular, T. Spencer has posed the question to prove that the $\ell^4$-norm of all eigenvectors are of order $N^{-1/4}$.

In this paper we prove several results in these directions for general Hermitian Wigner matrices. In Theorem 2.1 we give an upper bound on the eigenvalue density down to energy scales of order $\eta \geq \frac{\log N}{N}$.

Theorem 3.1 states that the density of states concentrates around its mean in probability sense down to energy windows of order $\eta \gg N^{-2/3}$ (modulo logarithmic corrections), improving the fluctuation result of [9] from scales $\eta \gg N^{-1/2}$. In Theorem 4.1 we prove that the expectation value of the density of states on scales $\eta \gg N^{-2/3}$ converges to the Wigner semicircle law. The previous best result [2] was valid for scales $\eta \gg N^{-1/2}$. These two theorems establish the validity of the Wigner semicircle law for all energy windows of order $\eta \gg N^{-2/3}$.

In Theorem 5.1 we show that most eigenvectors are fully extended in the sense that their $\ell^\infty$-norm is of order $N^{-1/2}$ (modulo logarithmic corrections).

We remark that this result can be easily obtained for all eigenvectors in the GUE case, by using the underlying unitary symmetry group. The reason why our proof of Theorem 5.1 does not apply to all eigenvectors is the lack of the lower bound on the density of states on the very short scales of $\eta \gg 1/N$. However, the results of Section 3 imply a bound of order $N^{2/3(1/p-1/2)}$ for the $\ell^p$-norm ($p \geq 2$) of *all* eigenvectors away from the spectral edge (Corollary 5.3).

In Theorem 6.1, by using the bounds on the eigenvectors, we give an estimate on the second moment of the Green function. Finally, in Theorem 7.1 we prove that no eigenvector is strongly localized in the sense that no eigenvector can be essentially supported on a small percentage of the sites. As a corollary, we show that the $\ell^p$-norm of the eigenvectors is $N^{1/p-1/2}$ for $1 \leq p < 2$.

We remark that all our results hold for the Wigner ensemble of real symmetric matrices as well. We will present the Hermitian case only, as the proofs for the real case require only obvious modifications.

In some of our results we need to assume further conditions on the distributions of the matrix elements in addition to (1.1). For convenience, we list the conditions we use in some of our theorems:

(C1) The function $g$ is twice differentiable and it satisfies

$$g''(x) \leq M,$$ (1.2)

with some finite $M$.

(C2) There exists a $\delta > 0$ such that

$$\int e^{\delta x^2} \, d\nu(x) < \infty,$$ (1.3)

and the same holds for $\widetilde{\nu}$.

(C3) The measures $\nu, \widetilde{\nu}$ satisfy the spectral gap inequality, that is, there exists a constant $C$ such that for any function $u$

$$\int \left| u - \int u \, d\nu \right|^2 d\nu \leq C \int |\nabla u|^2 \, d\nu,$$ (1.4)

and the same holds for $\widetilde{\nu}$.

(C4) The measures $\nu, \widetilde{\nu}$ satisfy the logarithmic Sobolev inequality, that is, there exists a constant $C$ such that for any density function $u > 0$ with $\int u \, d\nu = 1$,

$$\int u \log u \, d\nu \leq C \int |\nabla \sqrt{u}|^2 \, d\nu$$ (1.5)

and the same holds for $\widetilde{\nu}$.

We remark that (C4) implies (C3) and that all conditions are satisfied if $c_1 \leq g'', \widetilde{g}'' \leq c_2$ for some positive constants $c_1, c_2$.

NOTATION. We will use the notation $|A|$ both for the Lebesgue measure of a set $A \subset \mathbb{R}$ and for the cardinality of a discrete set $A \subset \mathbb{Z}$. The usual Hermitian scalar product for vectors $\mathbf{x}, \mathbf{y} \in \mathbb{C}^N$ will be denoted by $\mathbf{x} \cdot \mathbf{y}$ or by $(\mathbf{x}, \mathbf{y})$. We will use the convention that $C$ denotes generic large constants and $c$ denotes generic small positive constants whose values may change from line to line. Since we are interested in large matrices, we always assume that $N$ is sufficiently large.

## 2. Upper bound on the density of states.

The typical number of eigenvalues in an interval $I$ within the spectrum is expected to be of order $N|I|$. The following theorem proves the corresponding upper bound.

THEOREM 2.1. *Let $H$ be an $N \times N$ Wigner matrix as described in (1.1) and we assume condition (1.2). Let $I \subset \mathbb{R}$ be an interval with $|I| \geq (\log N)/N$ and denote by $\mathcal{N}_I$ the number of eigenvalues of $H$ in the interval $I$. Then there exists a constant $c > 0$ such that, for any $K$ large enough,*

$$(2.1) \qquad \mathbb{P}\{\mathcal{N}_I \geq KN|I|\} \leq e^{-cKN|I|}.$$

For a fixed spectral parameter, $z = E + i\eta$ with $E \in \mathbb{R}$, $\eta > 0$, we denote $G_z = (H - z)^{-1}$ the Green function. Let $\mu_1 \leq \mu_2 \leq \cdots \leq \mu_N$ be the eigenvalues of $H$ and let $F(x)$ be the empirical counting function of the eigenvalues

$$(2.2) \qquad F(x) = \frac{1}{N}|\{\alpha : \mu_\alpha \leq x\}|.$$

We define the Stieltjes transform of $F$ as

$$(2.3) \qquad m = m(z) = \frac{1}{N} \operatorname{Tr} G_z = \int_{\mathbb{R}} \frac{dF(x)}{x - z}$$

and we let

$$(2.4) \quad \rho = \rho_\eta(E) = \frac{\operatorname{Im} m(z)}{\pi} = \frac{1}{N\pi} \operatorname{Im} \operatorname{Tr} G_z = \frac{1}{N\pi} \sum_{\alpha=1}^{N} \frac{\eta}{(\mu_\alpha - E)^2 + \eta^2}$$

be the normalized density of states of $H$ around energy $E$ and regularized on scale $\eta$. The random variables $m$ and $\varrho$ also depend on $N$, and when necessary, we will indicate this fact by writing $m_N$ and $\varrho_N$.

The counting function $\mathcal{N}_I$ for intervals of length $|I| = \eta$ and the regularized density of states are closely related. On the one hand, for the interval $I = [E - \frac{\eta}{2}, E + \frac{\eta}{2}]$, we obviously have

$$(2.5) \qquad \mathcal{N}_I \leq CN|I|\varrho_\eta(E).$$

On the other hand, Theorem 2.1 provides the following upper bound for $m(z)$ under an additional assumption.

COROLLARY 2.2. *Let $z = E + i\eta$ with $E \in \mathbb{R}$ and $\eta \geq \log N/N$. We assume conditions (1.2) and (1.3). Then there exists $c > 0$ such that, for any sufficiently large $K$,*

$$(2.6) \qquad \mathbb{P}\left\{\sup_E |m(E + i\eta)| \leq K \log N\right\} \geq 1 - e^{-cKN\eta}.$$

*In particular, there exists a universal constant $C$ such that*

$$(2.7) \qquad \sup_E \mathbb{E}|m(E + i\eta)| \leq C \log N.$$

*The same bounds hold for the density without logarithmic factors*

$$\mathbb{P}\left\{\sup_E \varrho_\eta(E) \leq K\right\} \geq 1 - e^{-cKN\eta}, \qquad \sup_E \mathbb{E}\varrho_\eta(E) \leq C.$$

PROOF. It is well known that if the tail of the distribution of the matrix elements decays sufficiently fast, then the eigenvalues of $H$ lie within a compact set with the exception of an exponentially small probability. For completeness, we will prove in Lemma 7.4 that there is a universal constant $c_0$ depending only on $\delta$ in (1.3) such that, for any sufficiently large $K_0$, we have

$$(2.8) \qquad \mathbb{P}\left\{\max_\alpha |\mu_\alpha| \geq K_0\right\} \leq e^{-c_0 K_0^2 N}.$$

Cover the interval $[-K_0, K_0]$ by the union of subintervals $I_n = [(n - \frac{1}{2})\eta, (n + \frac{1}{2})\eta]$ of length $\eta$ where the integer index $n$ runs from $-[K_0\eta^{-1}] - 1$ to $[K_0\eta^{-1}] + 1$ (here $[\cdot]$ denotes the integer part). Clearly,

$$(2.9) \qquad |m(E + i\eta)| \leq \frac{\log N}{N\eta} \max_n \mathcal{N}_{I_n}, \qquad \varrho_\eta(E) \leq \frac{1}{N\eta} \max_n \mathcal{N}_{I_n},$$

assuming that $\max_\alpha |\mu_\alpha| \leq K_0$. Adding up the probabilities of the exceptional sets where $\mathcal{N}_{I_n} \geq K_0 N\eta$ and recalling $\eta \geq \log N/N$, we proved (2.6). The proof of (2.7) obviously follows from (2.6) and from the deterministic bounds $|m(E + i\eta)| \leq \eta^{-1}$. The bounds for the density $\varrho_\eta$ are proven similarly. This completes the proof of Corollary 2.2. □

In order to prove Theorem 2.1, we start with the following lemma:

LEMMA 2.3. *Suppose that $x_j$ and $y_j$, $j = 1, 2, \ldots, N$, are i.i.d. real random variables with mean zero and with a density function $(\text{const})e^{-g(x)}$. The expectation w.r.t. their joint probability measure $d\mu = (\text{const})\prod_{j=1}^N \times e^{-g(x_j) - g(y_j)} dx_j dy_j$ is denoted by $\mathbb{E}$. We assume that $g$ satisfies*

$$(2.10) \qquad g''(x) < M$$

*with some finite constant $M$. We set $z_j = x_j + \sqrt{-1}\,y_j$ and let $\mathbf{z} = (z_1, \ldots, z_N) \in \mathbb{C}^N$. Let $P$ be an orthogonal projection of rank $m$ in $\mathbb{C}^N$. Then for any constant $c > 0$ there exists a positive constant $\tilde{c}$, depending only on $c$ and $M$, such that*

$$\mathbb{E} \exp[-cX] \leq e^{-\tilde{c}m}, \qquad X = (P\mathbf{z}, P\mathbf{z}).$$

PROOF. Let $\mu_t$ be the probability measure on $\mathbb{R}^{2N} \cong \mathbb{C}^N$ given by

$$d\mu_t := Z_t^{-1} \exp[-tX]\, d\mu, \qquad Z_t = \int \exp[-tX]\, d\mu$$

and denote the expectation w.r.t. $\mu_t$ by $\mathbb{E}_t$. In case $t = 0$, we shall drop the subscript. The covariance of two random vectors $\mathbf{Y}, \mathbf{Z} \in \mathbb{C}^N$ w.r.t. the measure $\mu_t$ is denoted by

$$\langle \mathbf{Y}; \mathbf{Z} \rangle_{\mu_t} := \mathbb{E}_t(\mathbf{Y}, \mathbf{Z}) - (\mathbb{E}_t \mathbf{Y}, \mathbb{E}_t \mathbf{Z}).$$

Simple differentiation gives

$$\partial_t \log \mathbb{E} \exp[-tX] = -\mathbb{E}_t X = -\langle P\mathbf{z}; P\mathbf{z} \rangle_{\mu_t} - (\mathbb{E}_t P\mathbf{z}, \mathbb{E}_t P\mathbf{z}) \leq -\langle P\mathbf{z}; P\mathbf{z} \rangle_{\mu_t}.$$

Let $\nu_t$ denote the product measure on $\mathbb{R}^{2N} \cong \mathbb{C}^N$, with density for $z_j = x_j + \sqrt{-1}\,y_j$ to be proportional to $e^{-(M+2t)|z_j|^2/2}$, $j = 1, 2, \ldots, N$. We can rewrite

$$d\mu_t = Z_t^{-1} \exp[-tX]\, d\mu = \frac{d\mu_t}{d\nu_t}\, d\nu_t.$$

From the assumption (2.10) on $g$ and from $0 \leq P \leq I$, we obtain that $\frac{d\mu_t}{d\nu_t}$ is log convex on $\mathbb{R}^{2N}$. From the Brascamp–Lieb inequality (Theorem 5.4 in [5]) we have

$$\langle P\mathbf{z}; P\mathbf{z} \rangle_{\mu_t} \geq \langle P\mathbf{z}; P\mathbf{z} \rangle_{\nu_t}.$$

By computing the Gaussian covariance explicitly, there exists a constant $c' > 0$, depending only on $M$ and $c$, such that

$$\langle P\mathbf{z}; P\mathbf{z} \rangle_{\nu_t} \geq c'm \qquad \forall t \in [0, c].$$

We have thus obtained that

$$\partial_t \log \mathbb{E} \exp[-tX] \leq -c'm \qquad \forall t \in [0, c].$$

Integrating this inequality from $t = 0$ to $c$, we obtain the lemma. $\square$

REMARK. J. Bourgain [3] has informed us that the condition (2.10) can be removed.

We will use this result in the following setup. Let $\mathbf{v}_1, \mathbf{v}_2, \ldots, \mathbf{v}_{N-1}$ form an orthonormal basis in $\mathbb{C}^{N-1}$. Let

$$\xi_\alpha := |\mathbf{z} \cdot \mathbf{v}_\alpha|^2,$$

where the components of $\mathbf{z} = \mathbf{x} + \sqrt{-1}\mathbf{y} \in \mathbb{C}^{N-1}$ are distributed according to (const) $\prod_j e^{-g(x_j) - g(y_j)} \, dx_j \, dy_j$. With this notation, a standard large deviation argument yields the following corollary to Lemma 2.3:

COROLLARY 2.4. *Under the condition* (2.10), *there exists a positive $c$ such that, for any $\delta$ small enough,*

$$(2.11) \qquad \mathbb{P}\left( \sum_{\alpha \in A} \xi_\alpha \leq \delta m \right) \leq e^{-cm}$$

*for all $A \subset \{1, \ldots, N-1\}$ with cardinality $|A| = m$.*

PROOF OF THEOREM 2.1. To prove (2.1), we decompose the Hermitian $N \times N$ matrix $H$ as follows:

$$(2.12) \qquad H = \begin{pmatrix} h & \mathbf{a}^* \\ \mathbf{a} & B \end{pmatrix},$$

where $\mathbf{a} = (h_{12}, \ldots, h_{1N})^*$ and $B$ is the $(N-1) \times (N-1)$ matrix obtained by removing the first row and first column from $H$. Recall that $\mu_1 \leq \mu_2 \leq \cdots \leq \mu_N$ denote the eigenvalues of $H$ and let $\lambda_1 \leq \lambda_2 \leq \cdots \leq \lambda_{N-1}$ denote the eigenvalues of $B$. Note that $B$ is an $(N-1) \times (N-1)$ Hermitian Wigner matrix with a normalization off by a factor $(1 - \frac{1}{N})^{1/2}$. The following lemma is well known and we include a short proof for completeness.

LEMMA 2.5. (i) *With probability one, the eigenvalues of any Hermitian Wigner matrix* (1.1) *are simple.*
(ii) *The eigenvalues of $H$ and $B$ are interlaced:*

$$(2.13) \qquad \mu_1 < \lambda_1 < \mu_2 < \lambda_2 < \mu_3 < \cdots < \mu_{N-1} < \lambda_{N-1} < \mu_N.$$

PROOF. The proof of (i) follows directly from the continuity of the distribution of the matrix elements and is left to the reader. For the proof of (ii), suppose that $\mu$ is one of the eigenvalues of $H$. Let $\mathbf{v} = (v_1, \ldots, v_N)^t$ be a normalized eigenvector associated with $\mu$. From the continuity of the distribution it also follows that $v_1 \neq 0$ almost surely. From the eigenvalue equation $H\mathbf{v} = \mu\mathbf{v}$ and from (2.12) we find that

$$(2.14) \qquad hv_1 + \mathbf{a} \cdot \mathbf{w} = \mu v_1 \quad \text{and} \quad \mathbf{a}v_1 + B\mathbf{w} = \mu\mathbf{w},$$

with $\mathbf{w} = (v_2, \ldots, v_N)^t$. From these equations we obtain

$$\mathbf{w} = (\mu - B)^{-1}\mathbf{a}v_1 \quad \text{and, thus,}$$

(2.15)
$$(\mu - h)v_1 = \mathbf{a} \cdot (\mu - B)^{-1}\mathbf{a}v_1 = \frac{v_1}{N}\sum_\alpha \frac{\xi_\alpha}{\mu - \lambda_\alpha}$$

using the spectral representation of $B$, where we set

$$\xi_\alpha = |\sqrt{N}\mathbf{a} \cdot \mathbf{u}_\alpha|^2,$$

with $\mathbf{u}_\alpha$ being the normalized eigenvector of $B$ associated with the eigenvalue $\lambda_\alpha$. Since $v_1 \neq 0$, we have

(2.16)
$$\mu - h = \frac{1}{N}\sum_\alpha \frac{\xi_\alpha}{\mu - \lambda_\alpha},$$

where $\xi_\alpha$'s are strictly positive almost surely (notice that $\mathbf{a}$ and $\mathbf{u}_\alpha$ are independent). In particular, this shows that $\mu \neq \lambda_\alpha$ for any $\alpha$. In the open interval $\mu \in (\lambda_{\alpha-1}, \lambda_\alpha)$ the function

$$\Phi(\mu) := \frac{1}{N}\sum_\alpha \frac{\xi_\alpha}{\mu - \lambda_\alpha}$$

is strictly decreasing from $\infty$ to $-\infty$, therefore, there is exactly one solution to the equation $\mu - h = \Phi(\mu)$. Similar argument shows that there is also exactly one solution below $\lambda_1$ and above $\lambda_{N-1}$. This completes the proof. $\square$

We continue the proof of Theorem 2.1. Using the decomposition (2.12), we obtain the following formula for the Green function $G_z = (H - z)^{-1}$, $z = E + i\eta$ with $E \in \mathbb{R}$, $\eta > 0$:

(2.17) $\quad G_z(1,1) = \dfrac{1}{h - z - \mathbf{a} \cdot (B - z)^{-1}\mathbf{a}} = \left[ h - z - \dfrac{1}{N}\sum_{\alpha=1}^{N-1}\dfrac{\xi_\alpha}{\lambda_\alpha - z} \right]^{-1}.$

This formula in this context has already appeared in [1]. In particular, by considering only the imaginary part, we obtain

$$|G_z(1,1)| \leq \eta^{-1}\left| 1 + \frac{1}{N}\sum_{\alpha=1}^{N-1}\frac{\xi_\alpha}{(\lambda_\alpha - E)^2 + \eta^2} \right|^{-1}.$$

Similarly, for any $k = 1, 2, \ldots, N$, we define $B^{(k)}$ to be the $(N-1) \times (N-1)$ minor of $H$ obtained after removing the $k$th row and $k$th column. Let $\mathbf{a}^{(k)} = (h_{k1}, h_{k2}, \ldots, h_{k,k-1}, h_{k,k+1}, \ldots, h_{kN})^*$ be the $k$th column of $H$ without the $h_{kk}$ element. Let $\lambda_1^{(k)} < \lambda_2^{(k)} < \cdots$ be the eigenvalues and $\mathbf{u}_1^{(k)}, \mathbf{u}_2^{(k)}, \ldots$ the corresponding eigenvectors of $B^{(k)}$ and set $\xi_\alpha^{(k)} := N|\mathbf{a} \cdot \mathbf{u}_\alpha^{(k)}|^2$. Then we have the estimate

$$(2.18) \qquad |G_z(k,k)| \leq \eta^{-1} \left| 1 + \frac{1}{N} \sum_{\alpha=1}^{N-1} \frac{\xi_\alpha^{(k)}}{(\lambda_\alpha^{(k)} - E)^2 + \eta^2} \right|^{-1}.$$

For the interval $I \in \mathbb{R}$ given in Theorem 2.1, set $E$ to be its midpoint and $\eta = |I|$, that is, $I = [E - \frac{\eta}{2}, E + \frac{\eta}{2}]$. From (2.4), (2.5) and (2.18) we obtain

$$(2.19) \qquad \mathcal{N}_I \leq C\eta \sum_{k=1}^{N} |G_z(k,k)| \leq CN\eta^2 \sum_{k=1}^{N} \left| \sum_{\alpha : \lambda_\alpha^{(k)} \in I} \xi_\alpha^{(k)} \right|^{-1},$$

where we restricted the $\alpha$ summation in (2.18) only to eigenvalues lying in $I$.

For each $k = 1, 2, \ldots, N$, we define the event

$$\Omega_k := \left\{ \sum_{\alpha : \lambda_\alpha^{(k)} \in I} \xi_\alpha^{(k)} \leq \delta(\mathcal{N}_I - 1) \right\}$$

for some small $\delta > 0$. By the interlacing property of the $\mu_\alpha$ and $\lambda_\alpha^{(k)}$ eigenvalues, we know that there is at least $\mathcal{N}_I - 1$ eigenvalues of $B^{(k)}$ in $I$. By Corollary 2.4, there exists a positive universal constant $c$ such that $\mathbb{P}(\Omega_k) \leq e^{-c(\mathcal{N}_I - 1)}$. Setting $\widetilde{\Omega} = \bigcup_{k=1}^{N} \Omega_k$, we see that

$$(2.20) \qquad \mathbb{P}(\widetilde{\Omega} \text{ and } \mathcal{N}_I \geq KN|I|) \leq Ne^{-c(\mathcal{N}_I - 1)} \leq e^{-c'KN|I|}$$

if $K$ is sufficiently large, recalling that $\eta = |I| \geq \log N / N$. On the complement event, $\widetilde{\Omega}^c$, we have from (2.19) that

$$\mathcal{N}_I \leq \frac{CN^2\eta^2}{\delta(\mathcal{N}_I - 1)},$$

that is, $\mathcal{N}_I \leq (C/\delta)^{1/2}N\eta$. Choosing $K$ sufficiently large, we obtain (2.1) from (2.20). This proves Theorem 2.1. $\square$

## 3. Fluctuations of the density of states.

THEOREM 3.1. *Let $H$ be an $N \times N$ Wigner matrix as described in (1.1) and we assume the condition (1.2) and (1.3). Fix $E, \eta \in \mathbb{R}$ with $(\log N)/N \leq \eta \leq 1$ and set $z = E + i\eta$.*

(i) *Suppose that the measures $\nu, \widetilde{\nu}$ satisfy the spectral gap condition (1.4), then there exists a constant $C$ such that the covariance of the Stieltjes transform of the empirical eigenvalue distribution (2.3) satisfies*

$$\langle m(z); m(z) \rangle = \mathbb{E}|m(z) - \mathbb{E}m(z)|^2 \leq \frac{C}{N^2\eta^3}.$$

(ii) *Suppose that the measures $\nu$ and $\widetilde{\nu}$ satisfy the logarithmic Sobolev inequality (1.5), then there exists $c > 0$ such that*

$$(3.1) \qquad \mathbb{P}\{|m(z) - \mathbb{E}m(z)| \geq \varepsilon\} \leq e^{-cN\eta\varepsilon\min\{(\log N)^{-1}, N\eta^2\varepsilon\}}$$

*holds for any $\varepsilon > 0$.*

*The same bounds hold if $m(z)$ is replaced with the density of states $\varrho_\eta(E) = \frac{1}{\pi}\operatorname{Im}m(z)$.*

We remark that estimates on the covariance were obtained in [1, 2] down to scale $\eta \gg N^{-1/2}$. Concentration estimates down to the same scale were proven in [9].

PROOF. We start proving (i). Denote by $\mu_\alpha, \alpha = 1, \ldots, N$, the eigenvalues of $H$. Since, by the first order perturbation theory,

$$(3.2) \qquad \begin{aligned} \frac{\partial\mu_\alpha}{\partial\operatorname{Re}h_{ij}} &= \overline{\mathbf{v}}_\alpha(i)\mathbf{v}_\alpha(j) + \overline{\mathbf{v}}_\alpha(j)\mathbf{v}_\alpha(i) = 2\operatorname{Re}(\overline{\mathbf{v}}_\alpha(i)\mathbf{v}_\alpha(j)) \\ \frac{\partial\mu_\alpha}{\partial\operatorname{Im}h_{ij}} &= \sqrt{-1}[\overline{\mathbf{v}}_\alpha(i)\mathbf{v}_\alpha(j) - \overline{\mathbf{v}}_\alpha(j)\mathbf{v}_\alpha(i)] = 2\operatorname{Im}(\overline{\mathbf{v}}_\alpha(j)\mathbf{v}_\alpha(i)) \end{aligned}$$

for all $1 \leq i < j \leq N$ and

$$\frac{\partial\mu_\alpha}{\partial h_{ii}} = \overline{\mathbf{v}}_\alpha(i)\mathbf{v}_\alpha(i)$$

for all $i = 1, \ldots, N$, we obtain

$$\begin{aligned} &\langle m(z); m(z)\rangle \\ &\leq C\sum_{i<j}^N \mathbb{E}\left(\left|\frac{\partial m(z)}{\partial\sqrt{N}\operatorname{Re}h_{ij}}\right|^2 + \left|\frac{\partial m(z)}{\partial\sqrt{N}\operatorname{Im}h_{ij}}\right|^2\right) \\ &\quad + C\sum_{i=1}^N \mathbb{E}\left|\frac{\partial m(z)}{\partial\sqrt{N}h_{ii}}\right|^2 \\ &= \frac{C}{N^3}\sum_{i<j}^N \mathbb{E}\left(\left|\sum_\alpha \frac{1}{(\mu_\alpha - z)^2}\frac{\partial\mu_\alpha}{\partial\operatorname{Re}h_{ij}}\right|^2 + \left|\sum_\alpha \frac{1}{(\mu_\alpha - z)^2}\frac{\partial\mu_\alpha}{\partial\operatorname{Im}h_{ij}}\right|^2\right) \\ &\quad + \frac{C}{N^3}\sum_{i=1}^N \mathbb{E}\left|\sum_\alpha \frac{1}{(\mu_\alpha - z)^2}\frac{\partial\mu_\alpha}{\partial h_{ii}}\right|^2 \\ (3.3)\quad &= \frac{C}{N^3}\mathbb{E}\sum_{i<j}^N \sum_{\alpha,\beta} \frac{1}{(\mu_\alpha - z)^2}\frac{1}{(\mu_\beta - \bar{z})^2} \end{aligned}$$

$$\times \, [\mathrm{Re}(\overline{\mathbf{v}}_\alpha(i)\mathbf{v}_\alpha(j)) \, \mathrm{Re}(\overline{\mathbf{v}}_\beta(i)\mathbf{v}_\beta(j))$$
$$+ \, \mathrm{Im}(\overline{\mathbf{v}}_\alpha(j)\mathbf{u}_\alpha(i)) \, \mathrm{Im}(\overline{\mathbf{v}}_\beta(j)\mathbf{v}_\beta(i))]$$

$$+ \frac{C}{N^3}\mathbb{E}\sum_{i=1}^N\sum_{\alpha,\beta}\frac{1}{(\mu_\alpha-z)^2}\frac{1}{(\mu_\beta-\bar z)^2}|\mathbf{v}_\alpha(i)|^2|\mathbf{v}_\beta(i)|^2$$

$$= \frac{C}{N^3}\mathbb{E}\sum_{\alpha,\beta}\frac{1}{(\mu_\alpha-z)^2}\frac{1}{(\mu_\beta-\bar z)^2}\sum_{i,j}\mathbf{v}_\alpha(i)\overline{\mathbf{v}_\beta(i)}\mathbf{v}_\beta(j)\overline{\mathbf{v}_\alpha(j)}$$

$$= \frac{C}{N^3}\mathbb{E}\sum_\alpha\frac{1}{|\mu_\alpha-z|^4}.$$

Note that these identities hold without expectation as well. Now, for arbitrary $n \in \mathbb{Z}$, we define the interval

$$(3.4) \qquad\qquad I_n = [E + (n - \tfrac{1}{2})\eta; E + (n + \tfrac{1}{2})\eta].$$

Let $\mathcal{N}_{I_n} = |\{\alpha : \mu_\alpha \in I_n\}|$ denote the number of eigenvalues of $H$ in the interval $I_n$. For any $\eta \geq (\log N)/N$ it follows from Theorem 2.1 that

$$\mathbb{P}\{\mathcal{N}_{I_n} \geq KN\eta\} \leq e^{-cKN\eta}.$$

Therefore, for any fixed $K_0$ large enough, we find a constant $D$ such that $D\eta^{-1}$ is an integer and

$$
(3.5)\qquad
\begin{aligned}
\sum_\alpha \frac{1}{|\mu_\alpha-z|^4} &\leq \sum_{n=-D\eta^{-1}}^{D\eta^{-1}}\sum_{\alpha:\mu_\alpha\in I_n}\frac{1}{|\mu_\alpha-z|^4} + \sum_{\alpha:|\mu_\alpha|\geq K_0}\frac{1}{\eta^4}\\
&\leq \frac{C}{\eta^4}\sup_{|n|\leq D\eta^{-1}}\mathcal{N}_{I_n} + \frac{1}{\eta^4}|\{\alpha:|\mu_\alpha|\geq K_0\}|.
\end{aligned}
$$

From (3.3), we obtain

$$
(3.6)\qquad
\begin{aligned}
\langle m(z);m(z)\rangle &\leq \frac{C}{N^3\eta^4}\mathbb{E}\sup_{|n|\leq D\eta^{-1}}\mathcal{N}_{I_n} + \frac{1}{N^3\eta^4}\mathbb{E}|\{\alpha:|\mu_\alpha|\geq K_0\}|\\
&\leq \frac{CK}{N^2\eta^3} + \frac{C}{N^2\eta^4}\mathbb{P}\Big\{\sup_{|n|\leq D\eta^{-1}}\mathcal{N}_{I_n}\geq KN\eta\Big\}\\
&\quad + \frac{1}{N^2\eta^4}\mathbb{P}\{\exists\text{ eigenvalue }\mu\text{ with }|\mu|\geq K_0\}\\
&\leq \frac{CK}{N^2\eta^3} + \frac{CD}{N^2\eta^5}e^{-cKN\eta} + \frac{1}{N^2\eta^4}e^{-c_0K_0^2N},
\end{aligned}
$$

where we applied (2.1) for the second term with a sufficiently large $K$ and we used (2.8) in the third term to estimate the probability of finding an eigenvalue $|\mu| \geq K_0$. This proves part (i) of Theorem 3.1.

Next, we prove (ii). We will show how to control the real part of $m(z) - \mathbb{E}m(z)$, the imaginary part is controlled identically. Let $d\mathbb{P}$ denote the probability measure of the Hermitian Wigner matrix described in (1.1). Remark that

$$(3.7) \quad \frac{d}{d\beta}\left[e^{-\beta}\log\int\exp\left(e^\beta\operatorname{Re}[m(z) - \mathbb{E}m(z)]\right)d\mathbb{P}\right] = e^{-\beta}\int u\log u\,d\mathbb{P},$$

where we defined the probability density

$$u = \frac{\exp\left(e^\beta\operatorname{Re}[m(z) - \mathbb{E}m(z)]\right)}{\int\exp\left(e^\beta\operatorname{Re}[m(z) - \mathbb{E}m(z)]\right)d\mathbb{P}}.$$

From (3.7), we find, using the logarithmic Sobolev inequality and the bounds (3.3) and (3.5),

$$\frac{d}{d\beta}\left[e^{-\beta}\log\int\exp\left(e^\beta\operatorname{Re}[m(z) - \mathbb{E}m(z)]\right)d\mathbb{P}\right]$$

$$(3.8) \quad \leq Ce^{-\beta}\int|\nabla\sqrt{u}|^2\,d\mathbb{P}$$

$$\leq Ce^\beta\int\left\{\sum_{i<j}^N\left[\left|\frac{\partial m(z)}{\partial\sqrt{N}\operatorname{Re}h_{ij}}\right|^2 + \left|\frac{\partial m(z)}{\partial\sqrt{N}\operatorname{Im}h_{ij}}\right|^2\right] + \sum_{i=1}^N\left|\frac{\partial m(z)}{\partial\sqrt{N}h_{ii}}\right|^2\right\}u\,d\mathbb{P}$$

$$\leq \frac{Ce^\beta}{N^3\eta^4}\int u\sup_{|n|\leq D\eta^{-1}}\mathcal{N}_{I_n}\,d\mathbb{P} + \frac{Ce^\beta}{N^3\eta^4}\int|\{\alpha:|\mu_\alpha|\geq K_0\}|u\,d\mathbb{P}$$

$$\leq \frac{CKe^\beta}{N^2\eta^3} + \sum_{\ell\geq1}\frac{CKe^\beta(\ell+1)}{N^2\eta^3}$$

$$\times\int\mathbf{1}\left(K\ell\eta N\leq\sup_{|n|\leq D\eta^{-1}}\mathcal{N}_{I_n}\leq K(\ell+1)\eta N\right)$$

$$\times\mathbf{1}\left(\max_\alpha|\mu_\alpha|\leq K_0\right)u\,d\mathbb{P}$$

$$+ \frac{Ce^\beta}{N^3\eta^4}\|u\|_\infty\mathbb{P}\{\exists\alpha:|\mu_\alpha|\geq K_0\},$$

where we used the same intervals $I_n$ introduced in (3.4), and where the constants $K$, $D$ and $K_0$ have to be chosen sufficiently large.

To bound the second term on the r.h.s. of (3.8), we observe that, if $\sup_{|n|\leq D\eta^{-1}}\mathcal{N}_{I_n}\leq KN\eta(\ell+1)$ and if there is no $\alpha$ with $|\mu_\alpha|\geq K_0$, then, by using (2.9),

$$u\leq e^{e^\beta\operatorname{Re}[m(z) - \mathbb{E}m(z)]}\leq e^{2K\ell e^\beta\log N},$$

where we also used (2.7) and that $K$ is sufficiently large. Therefore, we obtain, for a large $K$,

$$\sum_{\ell \geq 1} \frac{CKe^{\beta}(\ell+1)}{N^2\eta^3} \int \mathbf{1}\Big(K\ell\eta N \leq \sup_{|n| \leq D\eta^{-1}} |\{\alpha : \mu_\alpha \in I_n\}| \leq K(\ell+1)\eta N\Big)$$

$$\times \mathbf{1}\Big(\max_\alpha |\mu_\alpha| \leq K_0\Big) u\, d\mathbb{P}$$

(3.9)

$$\leq \sum_{\ell \geq 1} \frac{CKe^{\beta}(\ell+1)}{N^2\eta^3} e^{2K\ell e^{\beta} \log N} \mathbb{P}\Big(\sup_{|n| \leq D\eta^{-1}} |\{\alpha : \mu_\alpha \in I_n\}| \geq K\ell\eta N\Big)$$

$$\leq \sum_{\ell \geq 1} \frac{CKe^{\beta}(\ell+1)}{N^2\eta^3} e^{-K\ell(c\eta N - 2e^{\beta} \log N)} \leq \frac{CKe^{\beta}}{N^2\eta^3},$$

as long as $e^{\beta} \leq \frac{cN\eta}{4\log N}$, where $c > 0$ is the constant from Theorem 2.1.

To bound the last term on the r.h.s. of (3.8), we use that $|m(z)| \leq \eta^{-1}$ and (2.8):

$$\frac{Ce^{\beta}}{N^3\eta^4} \|u\|_\infty \mathbb{P}\{\exists \alpha : |\mu_\alpha| \geq K_0\} \leq \frac{Ce^{\beta}}{N^3\eta^4} e^{C\eta^{-1}e^{\beta}} e^{-c_0 K_0^2 N} \leq \frac{Ce^{\beta}}{N^3\eta^4},$$

as long as $e^{\beta} \leq N\eta/C_0$ with a sufficiently big $C_0$.

Putting everything together, we obtain, from (3.8),

(3.10) $$\frac{d}{d\beta}\Big[e^{-\beta} \log \int \exp\big(e^{\beta} \operatorname{Re}[m(z) - \mathbb{E}m(z)]\big)\, d\mathbb{P}\Big] \leq \frac{Ce^{\beta}}{N^2\eta^3}$$

for all $\beta$ such that $e^{\beta} \leq \frac{N\eta}{C_1 \log N}$ with a sufficiently big $C_1$. Integrating this inequality from $\beta = \beta_0$ to $\beta = \log L$ with some $L \leq \frac{N\eta}{C_1 \log N}$, we find that

$$\log \mathbb{E} e^{L\operatorname{Re}[m(z)-\mathbb{E}m(z)]} \leq Le^{-\beta_0} \log \mathbb{E} \exp\big(e^{\beta_0} \operatorname{Re}[m(z) - \mathbb{E}m(z)]\big) + \frac{CL^2}{N^2\eta^3}.$$

Since $\mathbb{E}\operatorname{Re}[m(z) - \mathbb{E}m(z)] = 0$ and $|\operatorname{Re}[m(z) - \mathbb{E}m(z)]| \leq \eta^{-1}$ is uniformly bounded, by a second order Taylor expansion, we obtain that the first term on the r.h.s. vanishes as $\beta_0 \to -\infty$. Thus,

$$\mathbb{E} e^{L\operatorname{Re}[m(z)-\mathbb{E}m(z)]} \leq \exp\big(CL^2 N^{-2}\eta^{-3}\big).$$

Therefore,

(3.11) $$\mathbb{P}\{\operatorname{Re}[m(z) - \mathbb{E}m(z)] \geq \varepsilon\} \leq \exp\big(CL^2 N^{-2}\eta^{-3} - \varepsilon L\big)$$

$$\leq e^{-cN\eta\varepsilon \min\{(\log N)^{-1}, N\eta^2\varepsilon\}}$$

with a sufficiently small $c > 0$ after optimizing for $L$ under the condition $L \leq \frac{N\eta}{C_1 \log N}$. Replacing $m(z)$ with $-m(z)$ in the same proof, we obtain the estimate for $\mathbb{P}\{|\operatorname{Re}[m(z) - \mathbb{E}, m(z)]| \geq \varepsilon\}$.   $\square$

**4. Semicircle law on short scales.** For any $z = E + i\eta$, we let

$$m_{\mathrm{sc}} = m_{\mathrm{sc}}(z) = \int_{\mathbb{R}} \frac{\varrho_{\mathrm{sc}}(x)\, dx}{x - z}$$

be the Stieltjes transform of the Wigner semicircle distribution function whose density is given by

$$\varrho_{\mathrm{sc}}(x) = \frac{1}{2\pi} \sqrt{4 - x^2}\, \mathbf{1}(|x| \leq 2)\, dx.$$

For $\kappa, \widetilde{\eta} > 0$, we define the set

$$S_{N,\kappa,\widetilde{\eta}} := \{ z = E + i\eta \in \mathbb{C} : |E| \leq 2 - \kappa, \widetilde{\eta} \leq \eta \leq 1 \}$$

and for $\widetilde{\eta} = N^{-2/3} \log N$, we write

$$S_{N,\kappa} := \left\{ z = E + i\eta \in \mathbb{C} : |E| \leq 2 - \kappa, \frac{\log N}{N^{2/3}} \leq \eta \leq 1 \right\}.$$

THEOREM 4.1. *Let $H$ be an $N \times N$ Wigner matrix as described in (1.1) and assume the conditions (1.2), (1.3) and (1.5). Then for any $\kappa > 0$, the Stieltjes transform $m_N(z)$ [see (2.3)] of the empirical eigenvalue distribution of the $N \times N$ Wigner matrix satisfies*

$$(4.1) \qquad \lim_{N \to \infty} \sup_{z \in S_{N,\kappa}} |\mathbb{E} m_N(z) - m_{\mathrm{sc}}(z)| = 0.$$

Combining this result with Theorem 3.1, we obtain the following corollary:

COROLLARY 4.2. *Let $\kappa > 0$ and $\eta \in [N^{-2/3} \log N, 1]$ and assume the conditions of Theorem 4.1. Then we have*

$$(4.2) \qquad \mathbb{P}\left\{ \sup_{z \in S_{N,\kappa,\eta}} |m_N(z) - m_{\mathrm{sc}}(z)| \geq \varepsilon \right\} \leq e^{-cN\eta\varepsilon \min\{(\log N)^{-1}, N\eta^2 \varepsilon\}}$$

*for any $\varepsilon > 0$ and sufficiently large $N$. In particular, the density of states $\varrho_\eta(E)$ converges to the Wigner semicircle law in probability uniformly for all energies away from the spectral edges and for all energy windows at least $N^{-2/3} \log N$.*

*Let $\eta^* = \eta^*(N)$ such that $\eta \ll \eta^* \ll 1$ as $N \to \infty$, then we have the convergence of the counting function as well:*

$$(4.3) \qquad \mathbb{P}\left\{ \sup_{|E| \leq 2 - \kappa} \left| \frac{\mathcal{N}_{\eta^*}(E)}{2N\eta^*} - \varrho_{\mathrm{sc}}(E) \right| \geq \varepsilon \right\} \leq e^{-cN\eta\varepsilon \min\{(\log N)^{-1}, N\eta^2 \varepsilon\}}$$

*for any $\varepsilon > 0$ and sufficiently large $N$, where $\mathcal{N}_{\eta^*}(E) = |\{\alpha : |\mu_\alpha - E| \leq \eta^*\}|$ denotes the number of eigenvalues in the interval $[E - \eta^*, E + \eta^*]$.*

We remark that Bai et al. [2] have investigated the speed of convergence of the empirical eigenvalue distribution to the semicircle law. Their results directly imply (4.1) for $\eta = \operatorname{Im} z \gg N^{-1/2}$ and (4.3) for $\eta \geq N^{-2/5}$.

PROOF OF COROLLARY 4.2. For any two points $z, z' \in S_{N,\kappa,\eta}$, we have

$$|m_N(z) - m_N(z')| \leq C N^{4/3} |z - z_j|,$$

since the gradient of $m_N(z)$ is bounded by $C |\operatorname{Im} z|^{-2} \leq C N^{4/3}$ on $S_{N,\kappa,\eta}$. We can choose a set of at most $M = C \varepsilon^{-2} N^4$ points, $z_1, z_2, \ldots, z_M$, in $S_{N,\kappa,\eta}$ such that, for any $z \in S_{N,\kappa,\eta}$, there exists a point $z_j$ with $|z - z_j| \leq \varepsilon N^{-2}$. In particular, $|m_N(z) - m_N(z_j)| \leq \varepsilon/4$ if $N$ is large enough. Then using (3.1), we obtain

$$\mathbb{P}\left\{ \sup_{z \in S_{N,\kappa,\eta}} |m_N(z) - \mathbb{E} m_N(z)| \geq \varepsilon \right\} \leq \sum_{j=1}^{M} \mathbb{P}\left\{ |m_N(z_j) - \mathbb{E} m_N(z_j)| \geq \frac{\varepsilon}{2} \right\}$$

$$\leq e^{-cN\eta\varepsilon \min\{(\log N)^{-1}, N\eta^2\varepsilon\}}$$

under the condition that $\eta \geq N^{-2/3} \log N$ since $\operatorname{Im} z_j \geq \eta$. Combining this estimate with (4.1), we have proved (4.2).

To prove (4.3), we set

$$R(\lambda) = \frac{1}{\pi} \int_{E-M\eta}^{E+M\eta} \frac{\eta}{(\lambda - x)^2 + \eta^2} \, dx$$

$$= \frac{1}{\pi}\left[ \arctan\left( \frac{E-\lambda}{\eta} + M \right) - \arctan\left( \frac{E-\lambda}{\eta} - M \right) \right]$$

and let $\mathbf{1}_{I^*}(\lambda)$ denote the characteristic function of the interval $I^* = [E - \eta^*, E + \eta^*]$ with $\eta^* = M\eta$. From elementary calculus it follows that $\mathbf{1}_{I^*} - R$ can be decomposed into a sum of three functions, $\mathbf{1}_{I^*} - R = T_1 + T_2 + T_3$ with the following properties:

$$|T_1| \leq C M^{-1/2}, \qquad \operatorname{supp}(T_1) \in I_1 = [E - 2\eta^*, E + 2\eta^*];$$

$$|T_2| \leq 1, \qquad \operatorname{supp}(T_2) = J_1 \cup J_2,$$

where $J_1$ and $J_2$ are two intervals of length $M^{1/2}\eta$ with midpoint at $E - \eta^*$ and at $E + \eta^*$, respectively; and

$$|T_3(\lambda)| \leq \frac{C\eta\eta^*}{(\lambda - E)^2 + [\eta^*]^2}, \qquad \operatorname{supp}(T_3) \in I_1^c.$$

We thus have

$$
\begin{aligned}
(4.4) \quad \frac{\mathcal{N}_{\eta^*}(E)}{2N\eta^*} &= \frac{1}{2\eta^*} \int \mathbf{1}_{I^*}(\lambda) \, dF(\lambda) \\
&= \frac{1}{2\eta^*} \int R(\lambda) \, dF(\lambda) + \frac{1}{2\eta^*} \int [T_1(\lambda) + T_2(\lambda) + T_3(\lambda)] \, dF(\lambda).
\end{aligned}
$$

The last three terms are estimated trivially by

$$\frac{1}{2\eta^*} \int |T_1 + T_2 + T_3| \, dF$$

$$\leq \|T_1\|_\infty \frac{\mathcal{N}_{I_1}}{2N\eta^*} + \frac{\mathcal{N}_{J_1} + \mathcal{N}_{J_2}}{2N\eta^*} + \frac{C\eta}{\eta^*} \varrho_{\eta^*}(E)$$

$$\leq \frac{C}{M^{1/2}} [\varrho_{2\eta^*}(E) + \varrho_{M^{1/2}\eta}(E - \eta^*) + \varrho_{M^{1/2}\eta}(E + \eta^*) + \varrho_{\eta^*}(E)].$$

Using the bound (2.6), this error term is bounded by $CM^{-1/2}$ uniformly in $E$ apart from an event of exponentially small probability. In particular, this term is smaller than $\varepsilon/3$ if $M = \eta^*/\eta$ is sufficiently large.

The main term in (4.4) is computed as

$$\frac{1}{2\eta^*} \int R(\lambda) \, dF(\lambda) = \frac{1}{2\eta^*} \int_{E-\eta^*}^{E+\eta^*} \varrho_\eta(x) \, dx$$

$$= \frac{1}{2\eta^*} \int_{E-\eta^*}^{E+\eta^*} \varrho_{sc}(x) \, dx + \frac{1}{2\eta^*} \int_{E-\eta^*}^{E+\eta^*} [\varrho_\eta(x) - \varrho_{sc}(x)] \, dx$$

and the first term converges to $\varrho_{sc}(E)$ as long as $\eta^* \to 0$. Using (4.2), the second term is smaller than $\varepsilon/3$ apart from a set of probability $\exp(-cN\eta\varepsilon \times \min\{(\log N)^{-1}, N\eta^2\varepsilon\})$. Putting these estimates together, we arrive at (4.3). $\square$

PROOF OF THEOREM 4.1. Recall from the proof of Theorem 2.1 that $B^{(k)}$ denotes the $(N-1) \times (N-1)$ minor of $H$ after removing the $k$th row and $k$th column. Similarly to the definition of $m(z)$ in (2.3), we also define the Stieltjes transform of the density of states of $B^{(k)}$:

$$m^{(k)} = m^{(k)}(z) = \frac{1}{N-1} \text{Tr} \frac{1}{B^{(k)} - z} = \int_{\mathbb{R}} \frac{dF^{(k)}(x)}{x - z}$$

with the empirical counting function

$$F^{(k)}(x) = \frac{1}{N-1} |\{\alpha : \lambda_\alpha^{(k)} \leq x\}|,$$

where $\lambda_\alpha^{(k)}$ are the eigenvalues of $B^{(k)}$. The spectral parameter $z$ is fixed throughout the proof and we will omit from the argument of the Stieltjes transforms.

From a formula analogous to (2.17) but applied to the $k$th minor we get

$$(4.5) \quad m = \frac{1}{N} \sum_{k=1}^N G_z(k,k) = \frac{1}{N} \sum_{k=1}^N \frac{1}{h_{kk} - z - \mathbf{a}^{(k)} \cdot (1/(B^{(k)} - z))\mathbf{a}^{(k)}},$$

where recall that $\mathbf{a}^{(k)}$ is the $k$th column without the diagonal. Let $\mathbb{E}_k$ denote the expectation value w.r.t. the random vector $\mathbf{a}^{(k)}$. Define the random variable

$$(4.6) \qquad X_k := \mathbf{a}^{(k)} \cdot \frac{1}{B^{(k)} - z} \mathbf{a}^{(k)} - \mathbb{E}_k \mathbf{a}^{(k)} \cdot \frac{1}{B^{(k)} - z} \mathbf{a}^{(k)}$$

and note that

$$\mathbb{E}_k \mathbf{a}^{(k)} \cdot \frac{1}{B^{(k)} - z} \mathbf{a}^{(k)} = \frac{1}{N} \sum_\alpha \frac{1}{\lambda_\alpha^{(k)} - z} = \left(1 - \frac{1}{N}\right) m^{(k)}.$$

With this notation, it follows from (4.5) that

$$
\begin{aligned}
(4.7) \quad \mathbb{E}m = -\mathbb{E}\Big[ 1 \Big/ \Big\{ X_1 + \Big(1 - \frac{1}{N}\Big)[m^{(1)} - \mathbb{E}m^{(1)}] \\
+ \Big[\Big(1 - \frac{1}{N}\Big)\mathbb{E}m^{(1)} - \mathbb{E}m\Big] + [\mathbb{E}m + z] - h_{11} - \frac{1}{N}\mathbb{E}m^{(1)} \Big\} \Big],
\end{aligned}
$$

where we used that the distribution of $X_k$ and $m^{(k)}$ is independent of $k$.

Fix $\varepsilon > 0$. The first term in the denominator of (4.7) is estimated in the following lemma whose proof is given at the end of the section.

LEMMA 4.3. *For the random variable $X_1$ from (4.6), we have*

$$(4.8) \qquad \mathbb{E}|X_1|^4 \leq \frac{C(\log N)^2}{N^2 \eta^2},$$

*in particular,*

$$\mathbb{P}\{|X_1| \geq \varepsilon\} \leq \frac{C(\log N)^2}{N^2 \eta^2 \varepsilon^4}.$$

For the second term in the denominator in (4.7), we apply the large deviation estimate from Theorem 3.1, to the Stieltjes transform of $B^{(1)}$:

$$\mathbb{P}\{|m^{(1)} - \mathbb{E}m^{(1)}| \geq \varepsilon\} \leq e^{-cN\eta\varepsilon \min\{(\log N)^{-1}, N\eta^2\varepsilon\}}.$$

For the third term, we use that

$$
\begin{aligned}
\Big| m - \Big(1 - \frac{1}{N}\Big)m^{(1)} \Big| &= \Big| \int \frac{dF(x)}{x - z} - \Big(1 - \frac{1}{N}\Big) \int \frac{dF_1(x)}{x - z} \Big| \\
&= \frac{1}{N} \Big| \int \frac{NF(x) - (N-1)F_1(x)}{(x-z)^2} \, dx \Big|.
\end{aligned}
$$

By the interlacing property between the eigenvalues of $H$ and $B^{(1)}$, we have $\max_x |NF(x) - (N-1)F_1(x)| \leq 1$, thus,

$$\Big| m - \Big(1 - \frac{1}{N}\Big)m^{(1)} \Big| \leq \frac{1}{N} \int \frac{dx}{|x - z|^2} \leq \frac{C}{N\eta}$$

and, therefore, $|\mathbb{E}m - (1 - N^{-1})\mathbb{E}m^{(1)}| \leq C(N\eta)^{-1}$. Finally, from $\mathbb{E}x_{11}^2 < \infty$ we have

$$\mathbb{P}\{|h_{11}| \geq \varepsilon\} \leq \frac{C}{N\varepsilon^2}.$$

We define the set of events

$$\Omega := \{|X_1| \geq \varepsilon\} \cup \{|m^{(1)} - \mathbb{E}m^{(1)}| \geq \varepsilon\} \cup \{|h_{11}| \geq \varepsilon\},$$

then

$$\mathbb{P}(\Omega) \leq e^{-cN\eta\varepsilon\min\{1, N\eta^2\varepsilon\}} + \frac{C}{N\varepsilon^2} + \frac{C(\log N)^2}{N^2\eta^2\varepsilon^4}.$$

Let

$$Y = X_1 + (1 - N^{-1})[m^{(1)} - \mathbb{E}m^{(1)}] + [(1 - N^{-1})\mathbb{E}m^{(1)} - \mathbb{E}m] - h_{11},$$

then, similarly to (2.18), we have

$$|Y + \mathbb{E}m + z| \geq \left| \operatorname{Im} z + \mathbf{a}^{(k)} \cdot \frac{1}{B^{(k)} - z}\mathbf{a}^{(k)} \right| \geq \eta.$$

We also have $|\mathbb{E}m + z| \geq \eta$ since $\operatorname{Im} m \geq 0$. Set $\widetilde{Y} := Y \cdot \mathbf{1}_{\Omega^c}$, then obviously $|\widetilde{Y}| \leq 4\varepsilon$. Moreover, from (4.7) we have

$$(4.9) \quad \begin{aligned} &\mathbb{E}m + \frac{1}{\mathbb{E}m + z} \\ &= \mathbb{E}\mathbf{1}_{\Omega^c}\left[ \frac{1}{\mathbb{E}m + z} - \frac{1}{\mathbb{E}m + z + \widetilde{Y}} \right] + \mathbb{E}\mathbf{1}_{\Omega}\left[ \frac{1}{\mathbb{E}m + z} - \frac{1}{\mathbb{E}m + z + Y} \right]. \end{aligned}$$

The second term is bounded by

$$\left| \mathbb{E}\mathbf{1}_{\Omega}\left[ \frac{1}{\mathbb{E}m + z} - \frac{1}{\mathbb{E}m + z + Y} \right] \right| \leq 2\eta^{-1}\mathbb{P}(\Omega) \leq \frac{C}{\varepsilon^4 \log N} \leq C\varepsilon$$

uniformly for $z \in S_{N,\kappa}$ if $N \geq N(\varepsilon)$. In the first term we use the stronger bounds

$$|\mathbb{E}m + z| \geq \operatorname{Im} m(z) + \eta, \qquad |\mathbb{E}m + z + \widetilde{Y}| \geq \operatorname{Im} m(z) + \eta - 4\varepsilon$$

on the denominators. Thus, from (4.9) we obtain

$$(4.10) \quad \left| \mathbb{E}m + \frac{1}{\mathbb{E}m + z} \right| \leq \frac{C\varepsilon}{[\operatorname{Im} m(z) + \eta][\operatorname{Im} m(z) + \eta - 4\varepsilon]} + C\varepsilon$$

uniformly for $z \in S_{N,\kappa}$.

We note that the equation

$$(4.11) \quad M + \frac{1}{M + z} = 0$$

has a unique solution for any $z \in S_{N,\kappa}$ with $\mathrm{Im}\, M > 0$, namely, $M = m_{sc}(z)$, the Stieltjes transform of the semicircle law. Note that there exists $c(\kappa) > 0$ such that $\mathrm{Im}\, m_{sc}(E + i\eta) \geq c(\kappa)$ for any $|E| \leq 2 - \kappa$, uniformly in $\eta$.

The equation (4.11) is stable in the following sense. For any small $\delta$, let $M = M(z, \delta)$ be a solution to

$$(4.12) \qquad M + \frac{1}{M + z} = \delta,$$

with $\mathrm{Im}\, M > 0$. Subtracting (4.11) with $M = m_{sc}$ from (4.12), we have

$$(M - m_{sc})\left[ m_{sc} + z - \frac{1}{M + z} \right] = \delta(m_{sc} + z)$$

and

$$\mathrm{Im}\left[ m_{sc} + z - \frac{1}{M + z} \right] \geq \mathrm{Im}\, m_{sc} \geq c(\kappa).$$

Since the function $m_{sc} + z$ on the compact set $z \in S_{N,\kappa}$ is bounded, we get that

$$(4.13) \qquad |M - m_{sc}| \leq C_\kappa \delta$$

for some constant $C_\kappa$ depending only on $\kappa$.

Now we perform a continuity argument in $\eta$ to prove that

$$(4.14) \qquad |\mathbb{E}m(E + i\eta) - m_{sc}(E + i\eta)| \leq C^* \varepsilon$$

uniformly in $z \in S_{N,\kappa}$ with a sufficiently large constant $C^*$. Fix $E$ with $|E| \leq 2 - \kappa$. For $\eta = [\frac{1}{2}, 1]$, (4.14) follows from (4.10) with some small $\varepsilon$, since the right-hand side of (4.10) is bounded by $C\varepsilon$. Suppose now that (4.14) has been proven for some $\eta \in [2N^{-2/3} \log N, 1]$ and we want to prove it for $\eta/2$. By integrating the inequality

$$\frac{\eta/2}{(x - E)^2 + (\eta/2)^2} \geq \frac{1}{2} \frac{\eta}{(x - E)^2 + \eta^2}$$

with respect to $dF(x)$, we obtain that

$$\mathrm{Im}\, m\left( E + i\frac{\eta}{2} \right) \geq \frac{1}{2} \mathrm{Im}\, m(E + i\eta) \geq \frac{1}{2} c(\kappa) - C^* \varepsilon > \frac{c(\kappa)}{4}$$

for sufficiently small $\varepsilon$, where (4.14) and $\mathrm{Im}\, m_{sc}(E + i\eta) \geq c(\kappa)$ were used. Thus, the r.h.s. of (4.10) for $z = E + i\frac{\eta}{2}$ is bounded by $C\varepsilon$, the constant depending only on $\kappa$. Applying the stability bound (4.13), we get (4.14) for $\eta$ replaced with $\eta/2$. This completes the proof of Theorem 4.1. $\quad \square$

PROOF OF LEMMA 4.3. Recall that $\lambda_\alpha^{(1)}$ denote the eigenvalues and $\mathbf{u}_\alpha^{(1)}$ denote the eigenvectors of $B^{(1)}$ for $\alpha = 1, 2, \ldots, N - 1$. We also defined $\xi_\alpha^{(1)} =$ $|\mathbf{b}^{(1)} \cdot \mathbf{u}_\alpha(1)|^2$ with the vector $\mathbf{b}^{(1)} = (b_1, \ldots, b_{N-1}) = \sqrt{N}\mathbf{a}^{(1)} = \sqrt{N}(h_{12}, h_{13}, \ldots, h_{1N})^*$ whose components are i.i.d. random variables with real and imaginary parts distributed according to $\nu$. Dropping the sub- and superscripts, we have

$$X = \frac{1}{N}\sum_{\alpha=1}^{N-1} \frac{\xi_\alpha - 1}{\lambda_\alpha - z} = \frac{1}{N}\sum_\alpha \frac{\sum_{i,j} b_i \bar{b}_j \bar{\mathbf{u}}_\alpha(i)\mathbf{u}_\alpha(j) - 1}{\lambda_\alpha - z},$$

where all summations run from 1 to $N-1$.

Since the distribution $\nu$ satisfies the spectral gap inequality (1.4), we have

$$(4.15) \qquad \mathbb{E}|X|^2 \le C\mathbb{E}\sum_k \left[\left|\frac{\partial X}{\partial b_k}\right|^2 + \left|\frac{\partial X}{\partial \bar{b}_k}\right|^2\right],$$

where $\partial/\partial b = \frac{1}{2}[\partial/\partial(\mathrm{Re}\,b) - i\partial/\partial(\mathrm{Im}\,b)]$ and $\partial/\partial\bar{b} = \frac{1}{2}[\partial/\partial(\mathrm{Re}\,b) + i\partial/\partial(\mathrm{Im}\,b)]$. We compute

$$
\begin{aligned}
& \sum_k \left[\left|\frac{\partial X}{\partial b_k}\right|^2 + \left|\frac{\partial X}{\partial \bar{b}_k}\right|^2\right] \\
(4.16) \quad & = \sum_k \left[\left|\frac{1}{N}\sum_{\alpha,j} \frac{\bar{b}_j \bar{\mathbf{u}}_\alpha(k)\mathbf{u}_\alpha(j)}{\lambda_\alpha - z}\right|^2 + \left|\frac{1}{N}\sum_{\alpha,i} \frac{b_i \bar{\mathbf{u}}_\alpha(i)\mathbf{u}_\alpha(k)}{\lambda_\alpha - z}\right|^2\right] \\
& = \frac{1}{N^2}\sum_k \sum_{\alpha,\beta,i,j} \left[\frac{\bar{b}_j b_i \bar{\mathbf{u}}_\alpha(k)\mathbf{u}_\beta(k)\mathbf{u}_\alpha(j)\bar{\mathbf{u}}_\beta(i)}{(\lambda_\alpha - z)(\lambda_\beta - \bar{z})}\right. \\
& \qquad\qquad \left. + \frac{b_i \bar{b}_j \bar{\mathbf{u}}_\alpha(i)\mathbf{u}_\beta(j)\mathbf{u}_\alpha(k)\bar{\mathbf{u}}_\beta(k)}{(\lambda_\alpha - z)(\lambda_\beta - \bar{z})}\right] \\
& = \frac{2}{N^2}\sum_{\alpha,i,j} \frac{\bar{b}_j b_i \mathbf{u}_\alpha(j)\bar{\mathbf{u}}_\alpha(i)}{|\lambda_\alpha - z|^2}.
\end{aligned}
$$

Here we used the orthonormality of the eigenfunctions, $\sum_k \mathbf{u}_\alpha(k)\bar{\mathbf{u}}_\beta(k) = \delta_{\alpha,\beta}$. We insert this into (4.15) and take the expectation with respect to the $\mathbf{b}$ variables, $\mathbb{E}\bar{b}_j b_i = \delta_{ij}$, by using the fact that the components of $\mathbf{b}$ are independent of the $\lambda_\alpha$'s and $\mathbf{u}_\alpha$'s:

$$
\begin{aligned}
\mathbb{E}|X|^2 & \le \frac{C}{N^2}\mathbb{E}\sum_{\alpha,i,j} \frac{\bar{b}_j b_i \mathbf{u}_\alpha(j)\bar{\mathbf{u}}_\alpha(i)}{|\lambda_\alpha - z|^2} = \frac{C}{N^2}\mathbb{E}\sum_\alpha \frac{1}{|\lambda_\alpha - z|^2} \\
& \le \frac{C}{N\eta}\mathbb{E}\frac{1}{N}\sum_\alpha \frac{1}{|\lambda_\alpha - z|}.
\end{aligned}
$$

To estimate the last term, we have

$$
\begin{aligned}
(4.17) \quad \mathbb{E}\frac{1}{N}\sum_\alpha \frac{1}{|\lambda_\alpha - z|} & \le \int_{|\lambda| \le K_0} \frac{\mathbb{E}\varrho_\eta(\lambda)}{|\lambda - z|}\,d\lambda + \frac{1}{\eta}\mathbb{P}\{\max|\lambda_\alpha| \ge K_0\} \\
& \le C\log N.
\end{aligned}
$$

In the last step, by choosing $K_0$ sufficiently large, we used the uniform estimate (2.7) on $\mathbb{E}\varrho_\eta(\lambda)$ and the bound (2.8) for the eigenvalues of the $(N-1) \times (N-1)$ Wigner matrix $B^{(1)}$. Thus, we have showed that

$$(4.18) \qquad \mathbb{E}|X|^2 \le \frac{C \log N}{N\eta}.$$

To estimate the fourth moment, we have

$$\mathbb{E}|X|^4 = [\mathbb{E}|X|^2]^2 + \mathbb{E}[|X|^2 - \mathbb{E}|X|^2]^2$$
$$\le \frac{(C \log N)^2}{(N\eta)^2} + C\mathbb{E}\sum_k \left[ \left| \frac{\partial |X|^2}{\partial b_k} \right|^2 + \left| \frac{\partial |X|^2}{\partial \bar{b}_k} \right|^2 \right].$$

We will compute only the first term in the summation, the second one is identical. We have

$$C\mathbb{E}\sum_k \left| \frac{\partial |X|^2}{\partial b_k} \right|^2 \le 2C\mathbb{E}\left[ |X|^2 \sum_k \left( \left| \frac{\partial X}{\partial b_k} \right|^2 + \left| \frac{\partial X}{\partial \bar{b}_k} \right|^2 \right) \right]$$
$$\le \frac{1}{4}\mathbb{E}|X|^4 + C\mathbb{E}\left[ \sum_k \left( \left| \frac{\partial X}{\partial b_k} \right|^2 + \left| \frac{\partial X}{\partial \bar{b}_k} \right|^2 \right) \right]^2,$$

therefore,

$$(4.19) \qquad \frac{1}{2}\mathbb{E}|X|^4 \le \frac{(C \log N)^2}{(N\eta)^2} + C\mathbb{E}\left[ \sum_k \left( \left| \frac{\partial X}{\partial b_k} \right|^2 + \left| \frac{\partial X}{\partial \bar{b}_k} \right|^2 \right) \right]^2.$$

For the last term, we use (4.16):

$$\mathbb{E}\left[ \sum_k \left( \left| \frac{\partial X}{\partial b_k} \right|^2 + \left| \frac{\partial X}{\partial \bar{b}_k} \right|^2 \right) \right]^2$$
$$= \frac{1}{N^4}\mathbb{E}\left[ \sum_{\alpha,i,j} \frac{\bar{b}_j b_i \mathbf{u}_\alpha(j)\bar{\mathbf{u}}_\alpha(i)}{|\lambda_\alpha - z|^2} \right]^2$$
$$= \frac{1}{N^4}\mathbb{E}\sum_{\alpha,\beta}\sum_{i,j,\ell,m} \frac{\mathbb{E}[\bar{b}_j b_i \bar{b}_\ell b_m]\mathbf{u}_\alpha(j)\bar{\mathbf{u}}_\alpha(i)\mathbf{u}_\beta(\ell)\bar{\mathbf{u}}_\beta(m)}{|\lambda_\alpha - z|^2 |\lambda_\beta - z|^2}$$

$$(4.20) \quad = \frac{1}{N^4}\mathbb{E}\sum_{\alpha,\beta}\sum_{i\ne\ell} \frac{|\mathbf{u}_\alpha(i)|^2 |\mathbf{u}_\beta(\ell)|^2}{|\lambda_\alpha - z|^2 |\lambda_\beta - z|^2} + \frac{1}{N^4}\mathbb{E}\sum_{\alpha,\beta}\sum_{i\ne j} \frac{\mathbf{u}_\alpha(j)\bar{\mathbf{u}}_\alpha(i)\mathbf{u}_\beta(i)\bar{\mathbf{u}}_\beta(j)}{|\lambda_\alpha - z|^2 |\lambda_\beta - z|^2}$$
$$+ \frac{c_4}{N^4}\mathbb{E}\sum_{\alpha,\beta}\sum_i \frac{|\mathbf{u}_\alpha(i)|^2 |\mathbf{u}_\beta(i)|^2}{|\lambda_\alpha - z|^2 |\lambda_\beta - z|^2}$$
$$\le \frac{C}{N^4}\mathbb{E}\sum_{\alpha,\beta}\sum_{i,\ell} \frac{|\mathbf{u}_\alpha(i)|^2 |\mathbf{u}_\beta(\ell)|^2}{|\lambda_\alpha - z|^2 |\lambda_\beta - z|^2}$$

$$\leq \frac{C}{(N\eta)^2} \mathbb{E}\left[\frac{1}{N}\sum_\alpha \frac{1}{|\lambda_\alpha - z|}\right]^2.$$

In the second line we used that

$$\mathbb{E}[\bar{b}_j b_i \bar{b}_\ell b_m] = \delta_{ij}\delta_{\ell m}(1-\delta_{i\ell}) + \delta_{i\ell}\delta_{jm}(1-\delta_{im}) + c_4 \delta_{ij}\delta_{j\ell}\delta_{\ell m},$$

where $c_4 = \mathbb{E}|b|^4 = \int (x^2+y^2)^2 \, d\nu(x) \, d\nu(y)$. Finally, the last expectation value is estimated as

$$\mathbb{E}\left(\frac{1}{N}\sum_\alpha \frac{1}{|\lambda_\alpha - z|}\right)^2 \leq \mathbb{E}\left(\int_{|\lambda|\leq K_0} \frac{\varrho_\eta(\lambda)}{|\lambda - z|}\, d\lambda\right)^2 + \eta^{-2}\mathbb{P}\{\max|\lambda_\alpha| \geq K_0\}.$$

The second term is exponentially small by (2.8). In the first term we use (2.6) to conclude that $\varrho_\eta(\lambda) \leq K$ uniformly in $\lambda$, apart from an event of exponentially small probability. Inserting this bound into (4.20) and (4.19), we obtain the desired bound $\mathbb{E}|X|^4 \leq C(\log N)^2/(N\eta)^2$ in Lemma 4.3.  $\square$

## 5. Extended states.

Recall that the eigenvalues of $H$ are denoted by $\mu_1 < \mu_2 < \cdots < \mu_N$ and the corresponding normalized eigenvectors by $\mathbf{v}_1, \mathbf{v}_2, \ldots, \mathbf{v}_N$.

THEOREM 5.1. *Let $H$ be an $N \times N$ Wigner matrix as described in (1.1) and satisfying the conditions (1.2) and (1.3). Then there exist positive constants, $C_1, C_2$ and $c$, depending only on the constants $M$ in (1.2) and $\delta$ in (1.3), such that, for any $q > 0$,*

$$(5.1) \quad \mathbb{P}\left\{\frac{1}{N}\left|\left\{\beta : \max_j |\mathbf{v}_\beta(j)|^2 \geq \frac{C_1 q^2 (\log N)^2}{N}\right\}\right| \geq \frac{C_2}{q}\right\} \leq e^{-c(\log N)^2}.$$

REMARK. Suppose that $\|\mathbf{v}\|_\infty^2 \leq 1/L$ holds for an $\ell^2$-normalized vector $\mathbf{v} = (v_1, v_2, \ldots)$. Then the support of $\mathbf{v}$ contains at least $L$ elements. Thus, the quantity $\|\mathbf{v}\|_\infty^{-2}$ can be interpreted as the localization length of $\mathbf{v}$. With this interpretation, Theorem 5.1 states that the density of eigenstates with a localization length $L \leq Nq^{-2}$ (with logarithmic corrections) is bounded from above by $C/q$.

It also follows from Theorem 5.1 that, for every $p \geq 2$,

$$(5.2) \quad \mathbb{P}\left\{\frac{1}{N}|\{\beta : \|\mathbf{v}_\beta\|_{\ell^p} \geq C_1 N^{1/p-1/2}(\log N)^{2-4/p}\}| \geq \frac{C_2}{\log N}\right\} \leq e^{-c(\log N)^2}.$$

In other words, with high probability, all the $N$ eigenvectors, apart from a fraction converging to zero as $N \to \infty$, have the expected delocalization properties up to logarithmic corrections.

Note that, by duality, (5.2) immediately implies that

$$\mathbb{P}\left\{\frac{1}{N}|\{\beta : \|\mathbf{v}_\beta\|_{\ell^p} \leq C_1^{-1} N^{1/p-1/2}(\log N)^{2-4/p}\}| \geq \frac{C_2}{\log N}\right\} \leq e^{-c(\log N)^2}$$

(5.3)

for all $1 \leq p \leq 2$. In Section 7 we will improve (5.3) by showing, in Corollary 7.2, that, up to an event with exponentially small probability, every eigenvector $\mathbf{v}$ of $H$ satisfies $\|\mathbf{v}\|_p \leq cN^{1/p-1/2}$ for all $1 \leq p \leq 2$.

PROOF OF THEOREM 5.1. For brevity, we introduce the notation

$$\theta = [\log N]^2,$$

where $[\cdot]$ denotes the integer part. For $q > 0$, let $O_q$ denote the set of eigenvalue indices $\alpha$ such that the distance between the eigenvalues $\mu_{\alpha+\theta}$ and $\mu_{\alpha-\theta}$ is less than $q\theta/N$:

(5.4)
$$O_q = \left\{\alpha : |\mu_{\alpha-\theta} - \mu_{\alpha+\theta}| \leq \frac{q\theta}{N}\right\}.$$

Here we used the notation $\mu_\alpha = \mu_1$ if $\alpha < 1$ and $\mu_\alpha = \mu_N$ if $\alpha > N$. Given $K_0 > 0$, we define $\Omega$ to be the event characterized by all eigenvalues of $H$ being in the interval $[-K_0, K_0]$, that is,

$$\Omega = \{\omega : \sigma(H) \subset [-K_0, K_0]\}.$$

By (2.8), we have

$$\mathbb{P}(\Omega) \geq 1 - e^{-cN}$$

if $K_0$ is sufficiently large. We have

(5.5)
$$\mathbb{P}\left\{\frac{1}{N}\left|\left\{\beta : \max_j |\mathbf{v}_\beta(j)|^2 \geq \frac{C_1 q^2 \theta}{N}\right\}\right| \geq \frac{C_2}{q}\right\}$$

$$\leq \mathbb{P}\left\{\frac{1}{N}\left|\left\{\beta : \max_j |\mathbf{v}_\beta(j)|^2 \geq \frac{C_1 q^2 \theta}{N}\right\}\right| \geq \frac{C_2}{q} \text{ and } \Omega\right\} + e^{-cN}$$

$$\leq \mathbb{P}\left\{\frac{1}{N}\left|\left\{\beta : \max_j |\mathbf{v}_\beta(j)|^2 \geq \frac{C_1 q^2 \theta}{N}\right\} \cap O_q\right| \geq \frac{C_2}{2q} \text{ and } \Omega\right\}$$

$$+ \mathbb{P}\left\{|O_q^c| \geq \frac{C_2 N}{2q} \text{ and } \Omega\right\} + e^{-cN}.$$

A simple counting shows that the cardinality of the complement of $O_q$ is bounded by

$$|O_q^c| \leq \frac{CN}{q}$$

on $\Omega$. Therefore, by choosing $C_2$ sufficiently large, we have

$$\mathbb{P}\left\{\frac{1}{N}\left|\left\{\beta: \max_j|\mathbf{v}_\beta(j)|^2 \geq \frac{C_1 q^2\theta}{N}\right\}\right| \geq \frac{C_2}{q}\right\}$$

$$\leq \mathbb{P}\left\{\frac{1}{N}\left|\left\{\beta \in O_q: \max_j|\mathbf{v}_\beta(j)|^2 \geq \frac{C_1 q^2\theta}{N}\right\}\right| \geq \frac{C_2}{2q}\right\} + e^{-cN}$$

$$\leq \mathbb{P}\left\{\exists \beta \in O_q: \max_j|\mathbf{v}_\beta(j)|^2 \geq \frac{C_1 q^2\theta}{N}\right\} + e^{-cN}$$

$$\leq N \sup_\beta \mathbb{P}\left\{\beta \in O_q \text{ and } \max_j|\mathbf{v}_\beta(j)|^2 \geq \frac{C_1 q^2\theta}{N}\right\} + e^{-cN},$$

where we used that $q \ll N$ [if $q \geq N^{1/2}$, (5.1) is trivial]. The theorem now follows from Lemma 5.2 below. $\square$

LEMMA 5.2. *Under the assumptions of Theorem 5.1, there exists a constant $c > 0$ such that, for any sufficiently large $C$ and for any $q > 0$, we have*

$$(5.6) \qquad \sup_\beta \mathbb{P}\left\{\beta \in O_q \text{ and } \max_j|\mathbf{v}_\beta(j)|^2 \geq \frac{C\theta q^2}{N}\right\} \leq e^{-c\theta}.$$

PROOF. It is enough to prove that, for arbitrary $\beta \in \{1, \ldots, N\}$,

$$\mathbb{P}\left\{\beta \in O_q \text{ and } \max_j|\mathbf{v}_\beta(j)|^2 \geq \frac{C\theta q^2}{N}\right\} \leq e^{-c\theta}.$$

Therefore, we fix $\beta \in O_q$ and we consider first the $j = 1$ component $v_1 = \mathbf{v}_\beta(1)$ of $\mathbf{v}_\beta$; for brevity, we drop the index $\beta$ from the notation $\mu_\beta$ and $\mathbf{v}_\beta$. Set $\kappa := q\theta/N$. Recall that $\lambda_\alpha$ denotes the eigenvalues of $B$ in the decomposition (2.12). Denote by $A$ the set

$$A := \{\alpha : |\mu - \lambda_\alpha| \leq \kappa\}.$$

From the interlacing property of the eigenvalues, $|A| \geq \theta$ (if $\theta \leq \beta \leq N - \theta$, then actually $|A| = 2\theta$).

Recall the equations (2.14) and (2.15) obtained from the eigenvalue equation $H\mathbf{v} = \mu\mathbf{v}$ and from the decomposition (2.12). In particular, from (2.15) we find

$$(5.7) \qquad \|\mathbf{w}\|^2 = \mathbf{w} \cdot \mathbf{w} = |v_1|^2 \mathbf{a} \cdot (\mu - B)^{-2}\mathbf{a}.$$

Since $\|\mathbf{w}\|^2 = 1 - |v_1|^2$, we obtain

$$(5.8) \qquad |v_1|^2 = \frac{1}{1 + \mathbf{a} \cdot (\mu - B)^{-2}\mathbf{a}} = \frac{1}{1 + 1/N\sum_\alpha \xi_\alpha/(\mu - \lambda_\alpha)^2},$$

recalling the notation $\xi_\alpha = N|\mathbf{a} \cdot \mathbf{u}_\alpha|^2$, where $\mathbf{u}_\alpha$ is the normalized eigenvector of $B$ associated with the eigenvalue $\lambda_\alpha$. Thus, we have

$$(5.9) \qquad |v_1|^2 \leq \frac{1}{1 + N^{-1}\kappa^{-2}\sum_{\alpha \in A}\xi_\alpha} = \frac{N\kappa^2}{N\kappa^2 + \sum_{\alpha \in A}\xi_\alpha}.$$

Fix a small $\delta > 0$. Let $Q$ be the following event:

$$Q = \left\{\sum_{\alpha \in A}\xi_\alpha > \theta\delta\right\}.$$

On this set $Q$ we have the bound for $|v_1|^2$

$$\mathbf{1}_Q|v_1|^2 \leq \mathbf{1}_Q\frac{N\kappa^2}{N\kappa^2 + \sum_{\alpha \in A}\xi_\alpha} \leq \delta^{-1}N\kappa^2\theta^{-1} = \frac{\theta q^2}{N\delta}$$

and for $\delta$ small enough, we have

$$\mathbb{P}(Q^c) \leq e^{-c\theta}$$

by Corollary 2.4.

So far we have considered the $j = 1$ component of $\mathbf{v}$. We can repeat the argument for each $j = 1, 2, \ldots, N$. Thus, $Q$ should carry a subscript 1 and we can define $Q_j$ accordingly. Clearly, $\mathbb{P}\{(\bigcap_j Q_j)^c\} \leq Ne^{-c\theta} \leq e^{-c'\theta}$. On the other hand, on the set $\bigcap_j Q_j$ we have

$$\max_j |\mathbf{v}_\beta(j)|^2 \leq \frac{\theta q^2}{N\delta} \qquad \text{for any } \beta \in O_q. \quad \square$$

Theorem 5.1 implies that all eigenvectors of $H$, apart from a fraction vanishing in the limit $N \to \infty$, are completely extended, in the sense that, up to logarithmic corrections, $\|\mathbf{v}\|_\infty \leq \text{const}/N^{1/2}$. The reason we cannot prove this bound for all eigenvectors of $H$ is the lack of information about the microscopic distribution of the eigenvalues of $H$ (and of its minors) on scales of order $O(1/N)$. From Corollary 4.2, which gives precise information on the eigenvalue distribution up to scales of order $O(N^{-2/3}\log N)$, we can nevertheless get a nonoptimal bound on $\|\mathbf{v}\|_\infty$ for all eigenvectors of $H$.

PROPOSITION 5.3. *Let $H$ be an $N \times N$ Wigner matrix as described in (1.1) and satisfying the conditions (1.2), (1.3) and (1.5). Fix $\kappa > 0$, and assume that $C$ is large enough, depending on $\kappa$. Then there exists $c > 0$ such that*

$$\mathbb{P}\left\{\exists \ \mathbf{v} \ with \ H\mathbf{v} = \mu\mathbf{v}, \ \|\mathbf{v}\| = 1, \ \mu \in [-2 + \kappa, 2 - \kappa] \ and \ \|\mathbf{v}\|_\infty \geq \frac{C(\log N)}{N^{1/3}}\right\}$$

$$\leq e^{-c(\log N)^2}.$$

REMARK. The bound $\|\mathbf{v}\|_\infty \le CN^{-1/3}\log N$ obtained in this proposition trivially implies the upper bound $\|\mathbf{v}\|_p \le C(\log N)^{1-2/p}N^{2/3(1/p-1/2)}$ for $2 \le p < \infty$ as well.

PROOF OF PROPOSITION 5.3. Let $\eta^* = N^{-2/3}(\log N)^2$ and define

$$I_n = [-2 + \kappa + (n-1)\eta^*; -2 + \kappa + n\eta^*]$$

$$\text{for } n = 1, \ldots, n_{\max} = [(4-2\kappa)/\eta^*] + 1,$$

where $[x]$ denotes the integer part of $x \in \mathbb{R}$. Then

$$\bigcup_{n=1}^{n_{\max}} I_n \supset [-2 + \kappa, 2 - \kappa] \quad \text{and} \quad |I_n| = \eta^* = N^{-2/3}(\log N)^2$$

$$\text{for all } n = 1, \ldots, n_{\max}.$$

As before, let $\mathcal{N}_I = |\{\beta : \mu_\beta \in I\}|$ for any $I \subset \mathbb{R}$. Using (4.3) in Corollary 4.2, we have

$$\mathbb{P}\left\{\max_n \mathcal{N}_{I_n} \le \varepsilon N\eta^*\right\} \le e^{-c(\log N)^2}$$

if $\varepsilon$ is sufficiently small (depending on $\kappa$). Suppose that $\mu \in I_n$, and that $H\mathbf{v} = \mu\mathbf{v}$. From (5.8), we obtain

$$|v_1|^2 = \frac{1}{1 + 1/N\sum_\alpha \xi_\alpha/(\lambda_\alpha - \mu)^2} \le \frac{1}{1 + 1/4N\eta^2\sum_{\lambda_\alpha \in I_n}\xi_\alpha} \le \frac{4N\eta^2}{\sum_{\lambda_\alpha \in I_n}\xi_\alpha}$$

and from the interlacing property, there exist at least $\mathcal{N}_{I_n} - 1$ eigenvalues $\lambda_\alpha$ in $I_n$. Therefore,

$$\mathbb{P}\bigg(\exists\ \mathbf{v} \text{ with } H\mathbf{v} = \mu\mathbf{v},\ \|\mathbf{v}\| = 1,\ \mu \in [-2 + \kappa, 2 - \kappa] \text{ and } \|\mathbf{v}\|_\infty \ge \frac{C(\log N)}{N^{1/3}}\bigg)$$

$$\le \sum_{n=1}^{n_{\max}} \sum_{j=1}^{N} \mathbb{P}\bigg(\exists\ \mathbf{v} \text{ with } H\mathbf{v} = \mu\mathbf{v},\ \|\mathbf{v}\| = 1,\ \mu \in I_n \text{ and}$$

$$|v_j|^2 \ge \frac{C(\log N)^2}{N^{2/3}}\bigg)$$

$$\le Nn_{\max}\sup_n \mathbb{P}\bigg(\exists\ \mathbf{v} \text{ with } H\mathbf{v} = \mu\mathbf{v},\ \|\mathbf{v}\| = 1,\ \mu \in I_n \text{ and}$$

$$(5.10) \qquad\qquad\qquad |v_1|^2 \ge \frac{C(\log N)^2}{N^{2/3}}\bigg)$$

$$\le \text{const } N^{5/3}\sup_n \mathbb{P}\bigg(\sum_{\lambda_\alpha \in I_n}\xi_\alpha \le \frac{4N\eta^*}{C}\bigg)$$

$$\leq \text{const } N^{5/3} \sup_n \mathbb{P}\left( \sum_{\lambda_\alpha \in I_n} \xi_\alpha \leq \frac{4N\eta^*}{C} \text{ and } \mathcal{N}_{I_n} \geq \varepsilon N\eta^* \right)$$

$$+ \text{const } N^{5/3} \sup_n \mathbb{P}(\mathcal{N}_{I_n} \leq \varepsilon N\eta^*)$$

$$\leq \text{const } N^{5/3} e^{-cN^{1/3}} + \text{const } N^{5/3} e^{-c(\log N)^2} \leq e^{-c'(\log N)^2},$$

using Corollary 2.4 and choosing $C \geq 4(\delta\varepsilon)^{-1}$, where $\delta$ is from Corollary 2.4. □

**6. Second moment of the Green function.** In this section we use the result of Theorem 5.1 to obtain bounds on the second moment of the diagonal elements of the Green function of $H$. Recall the notation $\theta = [\log N]^2$.

THEOREM 6.1. *Let $H$ be an $N \times N$ Wigner matrix as described in (1.1) and satisfying the conditions (1.2) and (1.3). Let $z = E + i\eta$ be the spectral parameter of the Green function $G_{E,\eta} = G_z = (H - z)^{-1}$. Then there exist $c, C > 0$ such that, for any $\eta$,*

$$(6.1) \quad \mathbb{P}\left( \left| \left\{ E : \frac{1}{N} \sum_{j=1}^N |G_{E,\eta}(j,j)|^2 \geq C(\log N)^{12} \right\} \right| \geq \frac{C}{\log N} \right) \leq e^{-c(\log N)^2}.$$

REMARK. This theorem states that, with the exception of a very small probability, the second moment of the Green function, averaged over all sites, remains bounded (modulo logarithmic corrections) for all but a negligible set of energies in the sense of the Lebesgue measure.

PROOF. For any $k \in \mathbb{Z}$, we define the random sets

$$M_k := \left\{ \alpha : \frac{2^k}{N} < |\mu_{\alpha-\theta} - \mu_{\alpha+\theta}| \leq \frac{2^{k+1}}{N} \right\},$$

where we used again the convention that $\mu_\alpha = \mu_1$ for all $\alpha \leq 1$ and $\mu_\alpha = \mu_N$ for all $\alpha \geq N$. For given $\kappa, K_0 > 0$, let

$$\Omega_1 := \left\{ \bigcup_{k=0}^{\kappa \log N} M_k = \{1, 2, \ldots, N\} \right\} \cap \{\sigma(H) \subset [-K_0, K_0]\}.$$

From (2.8) we know that

$$\mathbb{P}\{\sigma(H) \in [-K_0, K_0]\} \geq 1 - e^{-cN},$$

for a sufficiently large $K_0$, so we obtain that $M_k = \varnothing$ for all $k \geq \kappa \log N$, if $\kappa$ is large enough, apart from an exponentially small event. From Theorem 2.1 we obtain that

$$\mathbb{P}\{M_k = \varnothing \text{ for all } k < 0\} \geq 1 - e^{-c\theta}$$

and, therefore, if $K_0$ and $\kappa$ are large enough,

$$\mathbb{P}(\Omega_1) \geq 1 - e^{-c\theta}.$$

In the sequel we will work on the set $\Omega_1$, that is, we can assume that the index $k$ runs from 0 to (const) $\log N$ and that all eigenvalues lie in $[-K_0, K_0]$.

By a simple counting, the cardinality of $M_k$ is bounded by

$$(6.2) \qquad |M_k| \leq (\text{const}) 2^{-k} N\theta$$

on the event $\Omega_1$.

For any $\alpha \in M_k$, denote

$$(6.3) \qquad \Omega_k(\alpha) := \left\{ \max_j |\mathbf{v}_\alpha(j)|^2 \leq C \frac{2^{2k}}{4N\theta} \right\},$$

where $\mathbf{v}_\alpha$ is the normalized eigenvector to the eigenvalue $\mu_\alpha$. From Lemma 5.2, we obtain, for every $k = 0, \ldots, \kappa \log N$,

$$\mathbb{P}\left\{ \bigcup_{\alpha \in M_k} \Omega_k^c(\alpha) \right\} \leq e^{-c\theta}$$

for some $c > 0$. Let

$$\Omega := \Omega_1 \cap \bigcap_k \bigcap_{\alpha \in M_k} \Omega_k(\alpha),$$

then

$$\mathbb{P}(\Omega) \geq 1 - e^{-c\theta}$$

for some $c > 0$. In the sequel we will work on the event $\Omega$.

Define the following random set of energies:

$$\mathcal{E} := \mathbb{R} \setminus \bigcup_k \bigcup_{\alpha \in M_k} \left\{ E : |\mu_\alpha - E| \leq \frac{2^k}{N\theta^2} \right\}.$$

The Lebesgue measure of the complement of $\mathcal{E}$ is bounded by

$$|\mathcal{E}^c| \leq \sum_k \sum_{\alpha \in M_k} \frac{2^{k+1}}{N\theta^2} \leq \frac{C}{\log N}.$$

Let $E \in \mathcal{E}$ and $\omega \in \Omega$. We compute

$$
\begin{aligned}
(6.4) \qquad \frac{1}{N} \sum_{j=1}^{N} |G_{E,\eta}(j,j)|^2 &\leq \frac{1}{N} \sum_{j=1}^{N} \sum_{k,\ell=0}^{\kappa \log N} \sum_{\alpha \in M_k} \sum_{\beta \in M_\ell} \frac{|\mathbf{v}_\alpha(j)|^2}{|\mu_\alpha - E|} \frac{|\mathbf{v}_\beta(j)|^2}{|\mu_\beta - E|} \\
&\leq \frac{2}{N} \sum_{j=1}^{N} \sum_{k \leq \ell}^{\kappa \log N} \sum_{\alpha \in M_k} \sum_{\beta \in M_\ell} \frac{|\mathbf{v}_\alpha(j)|^2}{|\mu_\alpha - E|} \frac{|\mathbf{v}_\beta(j)|^2}{|\mu_\beta - E|} \\
&\leq \frac{2}{N} \sum_{k \leq \ell}^{\kappa \log N} \frac{2^{2k} C}{4N\theta} \sum_{\alpha \in M_k} \sum_{\beta \in M_\ell} \frac{1}{|\mu_\alpha - E|} \frac{1}{|\mu_\beta - E|}.
\end{aligned}
$$

In the second line we used the symmetry between $\alpha$ and $\beta$, in the third line we used the estimate on $|\mathbf{v}_\alpha(j)|^2$ from (6.3) and that $\sum_j |\mathbf{v}_\beta(j)|^2 = 1$.

We now perform the $\alpha \in M_k$ summation; the $\beta \in M_\ell$ summation will be identical. Let $I$ be an arbitrary interval of length $|I| = 2^k/N$. We claim that the number of eigenvalues $\mu_\alpha \in I$ with $\alpha \in M_k$ is at most $2\theta$. We label the elements of $M_k$ in increasing order; $\alpha_1 < \alpha_2 < \cdots < \alpha_{|M_k|}$. Let $\mu_{\alpha_i}$ be the smallest eigenvalue in the set $I$ with index in $M_k$. If $i > |M_k| - 2\theta$, then there cannot be more than $2\theta$ eigenvalues with indices in $M_k$ in $I$. Otherwise, if $i \leq |M_k| - 2\theta$, we have

$$\mu_{\alpha_{i+2\theta}} - \mu_{\alpha_i} \geq \mu_{\alpha_{i+\theta}+\theta} - \mu_{\alpha_{i+\theta}-\theta} > \frac{2^k}{N}$$

and, therefore, since $|I| = 2^k/N$, $\mu_{\alpha_{i+2\theta}}$ cannot be in $I$.

We now define the intervals

$$I_m := \left[ E + \frac{2^k(m-1/2)}{N}, E + \frac{2^k(m+1/2)}{N} \right]$$

for each $m \in \mathbb{Z}$, $|m| \leq CN \cdot 2^{-k}$. Clearly, each $I_m$ contains at most $2\theta$ eigenvalues $\mu_\alpha$ with index $\alpha \in M_k$.

Notice that, for any $\mu \in I_m$, $m \neq 0$, we have $|\mu - E| \geq 2^{k-1}m/N$. For $\mu_\alpha \in I_0$, with $\alpha \in M_k$, by the choice of $E \in \mathcal{E}$, we have $|\mu_\alpha - E| \geq 2^k/(N\theta^3)$. Therefore,

$$
\begin{aligned}
\sum_{\alpha \in M_k} \frac{1}{|\mu_\alpha - E|} &\leq 2\theta \sum_{|m| \leq CN \cdot 2^{-k}} \max\left\{ \frac{1}{|\mu_\alpha - E|} : \alpha \in M_k, \mu_\alpha \in I_m \right\} \\
&\leq 2\theta \left[ \max\left\{ \frac{1}{|\mu_\alpha - E|} : \alpha \in M_k \right\} + 2 \sum_{m=1}^{CN \cdot 2^{-k}} \frac{N}{2^{k-1}m} \right] \\
&\leq \frac{C\theta^3 N}{2^k}.
\end{aligned}
$$
(6.5)

Using (6.5) both for the $\alpha$ and $\beta$ summations in (6.4), we obtain

$$\frac{1}{N} \sum_{j=1}^N |G_{E,\eta}(j,j)|^2 \leq \frac{2}{N} \sum_{k=0}^{\kappa \log N} \sum_{\ell=k}^{\kappa \log N} \frac{2^{2k}}{4N\theta} \frac{C\theta^3 N}{2^k} \frac{C\theta^3 N}{2^\ell} \leq C\theta^6$$

for any $E \in \mathcal{E}$ and $\omega \in \Omega$. This completes the proof of Theorem 6.1. $\quad\square$

## 7. Absence of localized eigenvectors.
In this section we show that eigenvectors of Wigner random matrices, up to events with exponentially small probability, cannot be localized in a strong sense given by the following definition.

DEFINITION 7.1. Let $L \geq 1$ be an integer and $\eta > 0$. We say that an $\ell^2$-normalized vector $\mathbf{v} = (v_1, \ldots, v_N) \in \mathbb{C}^N$ exhibits $(L, \eta)$-localization if there exists a set $A \subset \{1, 2, \ldots, N\}$ such that $|A| = L$ and $\sum_{j \in A^c} |v_j|^2 \leq \eta$.

THEOREM 7.1. *Let $H$ be an $N \times N$ Hermitian random matrix from the Wigner ensemble defined in (1.1), satisfying also the condition (1.2) and (1.3). Suppose that $\eta$ and $\nu = L/N$ are sufficiently small. Then, with a constant $c > 0$ that depends only on $M$ and $\delta$ from (1.2), (1.3), we have*

$$\mathbb{P}\{\exists \text{ a normalized eigenvector } \mathbf{v} \text{ of } H \text{ exhibiting } (L, \eta)\text{-localization}\} \leq e^{-cN}.$$

PROOF. Since, by (2.8),

$$\mathbb{P}\{\exists \text{ eigenvalue } \mu \text{ of } H \text{ with } |\mu| \geq K_0\} \leq e^{-cN}$$

if $K_0$ is large enough, it is sufficient to prove that

$$(7.1) \quad \sup_{\beta \in \{1, \ldots, N\}} \mathbb{P}\{\mathbf{v}_\beta \text{ exhibits } (L, \eta) \text{ localization and } |\mu_\beta| \leq K_0\} \leq e^{-cN},$$

where $\mu_1 \leq \mu_2 \leq \cdots \leq \mu_N$ denote the eigenvalues of $H$, and $\mathbf{v}_1, \mathbf{v}_2, \ldots, \mathbf{v}_N$ the corresponding normalized eigenvectors. To prove (7.1), we fix $\beta$, and consider the eigenvector $\mathbf{v}_\beta$ associated with the eigenvalue $\mu_\beta$; for brevity, we drop the index $\beta$ from $\mu_\beta$ and $\mathbf{v}_\beta$.

By the definition of $(L, \eta)$-localization and by the permutation symmetry

$$\mathbb{P}\{\mathbf{v} \text{ is } (L, \eta)\text{-localized and } |\mu| \leq K_0\}$$

$$(7.2) \quad = \mathbb{P}\left\{ \exists A \subset \{1, \ldots, N\} : |A| = L \text{ and } \sum_{j \in A^c} |v_j|^2 \leq \eta \text{ and } |\mu| \leq K_0 \right\}$$

$$\leq \binom{N}{L} \mathbb{P}\left\{ \sum_{j=L+1}^{N} |v_j|^2 \leq \eta \text{ and } |\mu| \leq K_0 \right\}.$$

We introduce the notation $\mathbf{u} = (v_1, \ldots, v_L)^t$, $\mathbf{w} = (v_{L+1}, \ldots, v_N)^t$ and for $j = L + 1, \ldots, N$,

$$\mathbf{c}_j = \frac{1}{\sqrt{N}} (h_{j1}, h_{j2}, \ldots, h_{jL})^* \in \mathbb{C}^L \quad \text{and}$$

$$\mathbf{d}_j = \frac{1}{\sqrt{N}} (h_{j,L+1}, \ldots, h_{jN})^* \in \mathbb{C}^{N-L}.$$

From the eigenvalue equation $H\mathbf{v} = \mu\mathbf{v}$, we obtain, for all $j \geq L + 1$,

$$\mu v_j = \mathbf{c}_j \cdot \mathbf{u} + \mathbf{d}_j \cdot \mathbf{w}$$

and, thus,

$$\sum_{j=L+1}^{N} |\mathbf{c}_j \cdot \mathbf{u}|^2 = \sum_{j=L+1}^{N} |\mu v_j - \mathbf{d}_j \cdot \mathbf{w}|^2 \leq 2\mu^2 \|\mathbf{w}\|^2 + 2 \sum_{j=L+1}^{N} |\mathbf{d}_j \cdot \mathbf{w}|^2.$$

Denoting by $D_1$ the $(N-L) \times L$ matrix with rows given by $\mathbf{c}_{L+1}^*, \dots, \mathbf{c}_N^*$ and by $D_2$ the $(N-L) \times (N-L)$ matrix with rows given by $\mathbf{d}_{L+1}^*, \dots, \mathbf{d}_N^*$, the last equation implies

$$(\mathbf{u}, D_1^* D_1 \mathbf{u}) \leq 2\mu^2 \|\mathbf{w}\|^2 + 2(\mathbf{w}, D_2^* D_2 \mathbf{w}) \leq 2\|\mathbf{w}\|^2 (\mu^2 + \lambda_{\max}(D_2^* D_2)).$$

Thus, from (7.2), we conclude that

$$\mathbb{P}\{\mathbf{v} \text{ is } (L, \eta)\text{-localized and } |\mu| \leq K_0\}$$

(7.3)
$$\leq \binom{N}{L} \mathbb{P}\{\|\mathbf{w}\|^2 \leq \eta \text{ and } |\mu| \leq K_0\}$$

$$\leq \binom{N}{L} \mathbb{P}\{(\mathbf{u}, D_1^* D_1 \mathbf{u}) \leq 2\eta(\mu^2 + \lambda_{\max}(D_2^* D_2)) \text{ and } |\mu| \leq K_0\}$$

$$\leq \binom{N}{L} \mathbb{P}\{(1-\eta)\lambda_{\min}(D_1^* D_1) \leq 2\eta(K_0^2 + \lambda_{\max}(D_2^* D_2))\}$$

$$\leq \binom{N}{L} \mathbb{P}\left\{ (1-\eta)\frac{N-L}{N} \lambda_{\min}(X_1^* X_1) \right.$$

$$\left. \leq 2\eta\left(K_0^2 + \frac{N-L}{N}\lambda_{\max}(X_2^* X_2)\right) \right\}$$

$$\leq \binom{N}{L} [\mathbb{P}\{\lambda_{\min}(X_1^* X_1) \leq c\} + \mathbb{P}\{\lambda_{\max}(X_2^* X_2) \geq C\}]$$

for any positive constants $c$ and $C$ if $\eta$ and $\nu = L/N$ are sufficiently small [because $(1-\eta)(1-\nu)c \geq 2\eta(K_0^2 + (1-\nu)C)$ if $\eta, \nu$ are sufficiently small]. Here $\lambda_{\min}(F)$ and $\lambda_{\max}(F)$ denotes the minimal and, respectively, the maximal eigenvalue of the Hermitian matrix $F$, and $X_1 = \sqrt{N/(N-L)} D_1$, $X_2 = \sqrt{N/(N-L)} D_2$. From Lemmas 7.3 and 7.4 below, we know that, for any sufficiently small $\nu = L/N$, for sufficiently large $C$, and for $c < 1/2$, there exists $\alpha > 0$ such that

(7.4)
$$\mathbb{P}\{\lambda_{\min}(X_1^* X_1) \leq c\} \leq e^{-\alpha(N-L)} \quad \text{and}$$

$$\mathbb{P}\{\lambda_{\max}(X_2^* X_2) \geq C\} \leq e^{-\alpha(N-L)}.$$

Thus, from (7.3), we obtain that, for $\eta > 0$ and $\nu = L/N$ small enough,

$$P\{\mathbf{v} \text{ is } (L, \eta)\text{-localized and } |\mu| \leq K_0\}$$

(7.5)
$$\leq 2 \binom{N}{L} e^{-\alpha(N-L)} \leq \left(\frac{e}{\nu}\right)^{\nu N} e^{-\alpha N(1-\nu)} \leq e^{-\alpha N/4}.$$

Since the constant $\alpha$ is independent of the eigenvalue $\mu$, (7.1) follows.  $\square$

COROLLARY 7.2.  *Suppose that the random matrix $H$ satisfies the same assumptions as in Theorem 7.1. Then, for every $\kappa > 0$ sufficiently small, there exists a constant $c > 0$ such that*

$$\mathbb{P}\{\exists \text{ normalized } \mathbf{v} \in \mathbb{C}^N \text{ such that } H\mathbf{v} = \mu\mathbf{v} \text{ and } \|\mathbf{v}\|_p \leq \kappa N^{1/p-1/2}\} \leq e^{-cN}$$

*for any $1 \leq p \leq 2$.*

REMARK.  If the eigenvector $\mathbf{v}$ is uniformly extended, that is, $|v_j|^2 = \frac{1}{N}$, then $\|\mathbf{v}\|_p = N^{1/p-1/2}$. This corollary indicates that the behavior of all eigenvectors is consistent with the extended states hypothesis as far as the low $\ell^p$-norms ($1 \leq p \leq 2$) are concerned.

PROOF OF COROLLARY 7.2.  From (2.8) with a sufficiently large $K_0$ we have

$$\mathbb{P}(\exists \text{ normalized } \mathbf{v} \in \mathbb{C}^N \text{ such that } H\mathbf{v} = \mu\mathbf{v} \text{ and } \|\mathbf{v}\|_p \leq \kappa N^{1/p-1/2})$$

(7.6)
$$\leq e^{-cK_0 N} + \mathbb{P}(\exists \text{ normalized } \mathbf{v} \in \mathbb{C}^N \text{ such that } H\mathbf{v} = \mu\mathbf{v},$$
$$|\mu| \leq K_0 \text{ and } \|\mathbf{v}\|_p \leq \kappa N^{1/p-1/2}).$$

Now, if $\mathbf{v}$ is a normalized eigenvector of $H$, associated with an eigenvalue $|\mu| \leq K_0$, we can apply Theorem 7.1. To this end, we fix $\nu$ and $\eta$ small enough, and let $L = \nu N$. After relabeling, we can assume that $|v_1| \geq |v_2| \geq \cdots \geq |v_L| \geq |v_{L+1}| \geq \cdots \geq |v_N|$. Then, by Theorem 7.1,

$$\mathbb{P}\left\{\sum_{j \leq L} |v_j|^2 \geq \eta\right\} \leq e^{-cN}.$$

Thus, with the exception of an event with exponentially small probability,

$$L|v_L|^2 \leq \sum_{j=1}^{L} |v_j|^2 \leq \eta.$$

This implies that $|v_L| \leq \sqrt{\eta/L}$. Therefore,

$$1 - \eta \leq \sum_{j \geq L+1} |v_j|^2 \leq |v_L|^{2-p} \sum_{j \geq L+1} |v_j|^p \leq (\eta/L)^{1-p/2} \sum_{j=1}^{N} |v_j|^p$$

and, hence,

$$\mathbb{P}\left(\sum_{j=1}^{N} |v_j|^p \leq L^{1-p/2}\frac{1-\eta}{\eta^{1-p/2}} = \kappa^p N^{1-p/2}\right) \leq e^{-cN},$$

which, together with (7.6), completes the proof. $\square$

In the next two lemmas we prove effective the large deviation estimate on the largest and the smallest eigenvalue of some covariance matrices.

LEMMA 7.3.   *Let $X = (X_{ij})$ be a complex $N \times L$ matrix, with $N > L$, such that, for all $i = 1, \ldots, N$ and $j = 1, \ldots, L$, $\mathrm{Re}\, X_{ij}, \mathrm{Im}\, X_{ij}$ are i.i.d. random variables with*

$$\mathbb{E} X_{ij} = 0, \qquad \mathbb{E}|X_{ij}|^2 = \frac{1}{2N} \quad and \quad \mathbb{E} e^{\delta N|X_{ij}|^2} \leq K_\delta < \infty$$

*for some $\delta > 0$ and with $K_\delta$ independent of $N$:*

(i) *For $C > 0$ large enough,*

$$\mathbb{P}(\lambda_{\max}(X^*X) \geq C) \leq e^{-c_0 CN}$$

*for a constant $c_0$ depending only on $\delta$.*

(ii) *For $\nu = L/N$ sufficiently small and for all $0 < c < 1/2$, there exists $\alpha_0 = \alpha_0(\delta, c, \nu) > 0$ such that*

$$\mathbb{P}(\lambda_{\min}(X^*X) \leq c) \leq e^{-\alpha N}$$

*for all $\alpha < \alpha_0$.*

REMARK.   The precise large deviation rate function for $\lambda_{\min}$ and $\lambda_{\max}$ was determined recently in [7] in the limit $N \to \infty$ under the additional condition that $L = o(N/\log\log N)$. Our proof is somewhat different and it also applies to the case $L \leq \nu N$, with $\nu$ small enough, but the decay rate we obtain is not precise. The history and earlier results in this direction were reviewed in [7] and we shall not repeat it here.

PROOF.   We begin by proving (i). First, fix $\mathbf{z} \in \mathbb{C}^L$, with $\|\mathbf{z}\| = 1$. We claim that

(7.7)          $$\mathbb{P}\{(\mathbf{z}, X^*X\mathbf{z}) \geq C\} \leq e^{-c_1 CN}$$

for a constant $c_1$ depending only on $\delta$. In fact, for arbitrary $\kappa > 0$,

(7.8)
$$\mathbb{P}\{(\mathbf{z}, X^*X\mathbf{z}) \geq C\} \leq e^{-\kappa CN}\mathbb{E} e^{\kappa N(\mathbf{z}, X^*X\mathbf{z})}$$
$$= e^{-\kappa CN}\mathbb{E} e^{\kappa N \sum_{j=1}^{N} |\mathbf{X}_j \cdot \mathbf{z}|^2},$$

where, for $j = 1, \ldots, N$, $\mathbf{X}_j = (X_{j1}, \ldots, X_{jL})^*$ denotes the adjoint of the $j$th row of $X$. Since different rows of $X$ are independent and identically distributed, we find

$$(7.9) \quad \mathbb{P}\{(\mathbf{z}, X^* X \mathbf{z}) \geq C\} \leq e^{-\kappa CN} \prod_{j=1}^{N} \mathbb{E} e^{\kappa N |\mathbf{X}_j \cdot \mathbf{z}|^2} = e^{-\kappa CN} (\mathbb{E} e^{\kappa N |\mathbf{X}_1 \cdot \mathbf{z}|^2})^N.$$

Consider now the random vector $\mathbf{Y} = \sqrt{N} \mathbf{X}_1 = (y_1, \ldots, y_L)^*$ with i.i.d. components. We have

$$
\begin{aligned}
\mathbb{E} e^{\kappa |\mathbf{Y} \cdot \mathbf{z}|^2} &= \mathrm{const} \int_{\mathbb{R} \times \mathbb{R}} dq \, dp \, e^{-(q^2 + p^2)/4} \mathbb{E} e^{\sqrt{\kappa}(q \operatorname{Re}(\mathbf{Y} \cdot \mathbf{z}) + p \operatorname{Im}(\mathbf{Y} \cdot \mathbf{z}))} \\
&= \mathrm{const} \int_{\mathbb{R} \times \mathbb{R}} dq \, dp \, e^{-(q^2 + p^2)/4} \prod_{i=1}^{L} \mathbb{E} e^{\sqrt{\kappa}(q \operatorname{Re}(z_i y_i) + p \operatorname{Im}(z_i y_i))} \\
&= \mathrm{const} \int_{\mathbb{R} \times \mathbb{R}} dq \, dp \, e^{-(q^2 + p^2)/4} \\
&\qquad \times \prod_{i=1}^{L} \mathbb{E} e^{\sqrt{\kappa}(q \operatorname{Re} z_i + p \operatorname{Im} z_i) \operatorname{Re} y_i} \mathbb{E} e^{\sqrt{\kappa}(-q \operatorname{Im} z_i + p \operatorname{Re} z_i) \operatorname{Im} y_i}
\end{aligned}
$$
(7.10)

with an appropriate normalization constant. Since $\mathbb{E} \operatorname{Re} y_i = 0$, we find, for arbitrary $r \in \mathbb{R}$,

$$
\begin{aligned}
\mathbb{E} e^{r \operatorname{Re} y_i} &= \sum_{n \geq 0} \frac{r^n}{n!} \mathbb{E}(\operatorname{Re} y_i)^n \\
&= 1 + \sum_{n \geq 1} \frac{r^{2n}}{(2n)!} \mathbb{E}(\operatorname{Re} y_i)^{2n} + \sum_{n \geq 1} \frac{r^{2n+1}}{(2n+1)!} \mathbb{E}(\operatorname{Re} y_i)^{2n+1}.
\end{aligned}
$$
(7.11)

Using that, for all $n \geq 1$,

$$\frac{r^{2n+1}}{(2n+1)!} \mathbb{E}(\operatorname{Re} y_i)^{2n+1} \leq \frac{r^{2n}}{(2n)!} \mathbb{E}(\operatorname{Re} y_i)^{2n} + \frac{r^{2n+2}}{(2n+2)!} \mathbb{E}(\operatorname{Re} y_i)^{2n+2},$$

we obtain that

$$
\begin{aligned}
\mathbb{E} e^{r \operatorname{Re} y_i} &= \sum_{n \geq 0} \frac{r^n}{n!} \mathbb{E}(\operatorname{Re} y_i)^n = 1 + 3 \sum_{n \geq 1} \frac{r^{2n}}{(2n)!} \mathbb{E}(\operatorname{Re} y_i)^{2n} \\
&\leq 1 + \sum_{n \geq 1} \frac{n!(3r)^{2n}}{\delta^{2n}(2n)!} \mathbb{E} e^{\delta(\operatorname{Re} y_i)^2} \\
&\leq 1 + \sum_{n \geq 1} \frac{(3r)^{2n} K_\delta^n}{n! \delta^{2n}} \leq e^{9 K_\delta r^2 / \delta^2},
\end{aligned}
$$
(7.12)

where we chose $\delta > 0$ small enough, and we used that $K_\delta = \mathbb{E}e^{\delta y^2} = \int e^{\delta y^2} e^{-g(y)} \, dy < \infty$. Since $\|\mathbf{z}\| = 1$, from (7.10) we obtain

$$(7.13) \quad \mathbb{E}e^{\kappa|\mathbf{Y}\cdot\mathbf{z}|^2} \leq \text{const} \int_{\mathbb{R}\times\mathbb{R}} dq \, dp \, e^{-(q^2+p^2)(1/4-36\kappa(K_\delta/\delta^2))} \leq \text{const}$$

by choosing $\kappa > 0$ small enough. Inserting in (7.9) and choosing $C$ large enough, we find (7.7).

Now, for fixed $0 < \varepsilon < 1/4$, we choose a family $\{\mathbf{z}_j\}_{j\in I}$ with $\mathbf{z}_j \in \mathbb{C}^L$, $\|\mathbf{z}_j\| \leq 1$ for all $j \in I$, such that $|I| \leq (2/\varepsilon)^{2L}$, and such that, for all $\mathbf{z} \in \mathbb{C}^L$ with $\|\mathbf{z}\| = 1$, there exists $j \in I$ with $\|\mathbf{z} - \mathbf{z}_j\| \leq \varepsilon$. For a suitable $j \in I$, we have

$$\|X^*X\| = \sup_{\mathbf{z}\in\mathbb{C}^N} (\mathbf{z}, X^*X\mathbf{z}) = (\mathbf{z}_{\max}, X^*X\mathbf{z}_{\max})$$

$$(7.14)$$

$$\leq 2\|\mathbf{z}_{\max} - \mathbf{z}_j\|\|X^*X\| + (\mathbf{z}_j, X^*X\mathbf{z}_j) \leq 2\varepsilon\|X^*X\| + (\mathbf{z}_j, X^*X\mathbf{z}_j)$$

and, thus, if $\lambda_{\max}(X^*X) \geq C$, there must be at least one $j \in I$ such that

$$(\mathbf{z}_j, X^*X\mathbf{z}_j) \geq (1 - 2\varepsilon)C.$$

Therefore, since $|I| \leq (2/\varepsilon)^{2L}$, we can apply (7.7) to obtain

$$\mathbb{P}\{\lambda_{\max}(X^*X) \geq C\} \leq \mathbb{P}\{\exists j \in I : (\mathbf{z}_j, X^*X\mathbf{z}_j) \geq (1 - 2\varepsilon)C\}$$

$$(7.15) \qquad\qquad \leq (2/\varepsilon)^{2L} \sup_j \mathbb{P}\{(\mathbf{z}_j, X^*X\mathbf{z}_j) \geq (1 - 2\varepsilon)C\}$$

$$\leq (2/\varepsilon)^{2L} e^{-c_1 CN}$$

and, thus, for $C$ large enough (and since $L \leq N$),

$$\mathbb{P}\{\lambda_{\max}(X^*X) \geq C\} \leq e^{-(c_1/2)CN}.$$

Next, we prove (ii). Again, we first fix $\mathbf{z} \in \mathbb{C}^L$, and prove that, for $0 < c < 1/2$, and for all $\alpha$ sufficiently small,

$$(7.16) \qquad\qquad \mathbb{P}\{(\mathbf{z}, X^*X\mathbf{z}) \leq c\} \leq e^{-\alpha N}.$$

To this end, we observe that, for $\beta > 0$,

$$(7.17) \quad \mathbb{P}\{(\mathbf{z}, X^*X\mathbf{z}) \leq c\} \leq e^{\beta Nc}\mathbb{E}e^{-\beta N(\mathbf{z}, X^*X\mathbf{z})} = (e^{\beta c}\mathbb{E}e^{-\beta|\mathbf{Y}\cdot\mathbf{z}|^2})^N,$$

where we defined, as before, $\mathbf{Y} = \sqrt{N}\mathbf{X}_1 = (y_1, \ldots, y_L)^*$. Since $e^{-\beta r} \leq 1 - \beta r + \beta^2 r^2/2$ for all $r \geq 0$, we obtain

$$\mathbb{E}e^{-\beta|\mathbf{Y}\cdot\mathbf{z}|^2} \leq 1 - \beta\mathbb{E}|\mathbf{Y}\cdot\mathbf{z}|^2 + \frac{\beta^2}{2}\mathbb{E}|\mathbf{Y}\cdot\mathbf{z}|^4 = 1 - \frac{\beta}{2} + O(\beta^2) \leq e^{-\beta/2+O(\beta^2)}$$

if $\beta > 0$ is sufficiently small depending only on $\mathbb{E}y_1^4$. Therefore, we find

$$e^{\beta c}\mathbb{E}e^{-\beta|\mathbf{Y}\cdot\mathbf{z}|^2} \leq e^{-\beta(1/2-c)+O(\beta^2)},$$

which proves (7.16) from (7.17) with a sufficiently small $\alpha$, depending on $c$.

To conclude the proof of (ii), we fix $\varepsilon > 0$ and a family $\{\mathbf{z}_j\}_{j \in I}$ with $\mathbf{z}_j \in \mathbb{C}^L$, $\|\mathbf{z}_j\| \leq 1$ for all $j \in I$, such that, for all $\mathbf{z} \in \mathbb{C}^L$ with $\|\mathbf{z}\| = 1$, there exists $j \in I$ with $\|\mathbf{z} - \mathbf{z}_j\| \leq \varepsilon$ and $|I| \leq (2/\varepsilon)^{2L}$. Then, for a suitable $j \in I$,

$$
\begin{aligned}
\lambda_{\min}(X^*X) &= \inf_{\|\mathbf{z}\|=1} (\mathbf{z}, X^*X\mathbf{z}) = (\mathbf{z}_{\min}, X^*X\mathbf{z}_{\min}) \\
(7.18) \qquad &\geq (\mathbf{z}_j, X^*X\mathbf{z}_j) - 2\|\mathbf{z}_{\min} - \mathbf{z}_j\|\lambda_{\max}(X^*X) \\
&\geq (\mathbf{z}_j, X^*X\mathbf{z}_j) - 2\varepsilon\lambda_{\max}(X^*X).
\end{aligned}
$$

Therefore, we find

$$
\begin{aligned}
&\mathbb{P}\{\lambda_{\min}(X^*X) \leq c\} \\
(7.19) \qquad &\leq \mathbb{P}\{\lambda_{\min}(X^*X) \leq c \text{ and } \lambda_{\max}(X^*X) \leq C\} + \mathbb{P}\{\lambda_{\max}(X^*X) \geq C\} \\
&\leq \mathbb{P}\{\exists j : (\mathbf{z}_j, X^*X\mathbf{z}_j) \leq c + 2\varepsilon C\} + \mathbb{P}\{\lambda_{\max}(X^*X) \geq C\} \\
&\leq \left(\frac{2}{\varepsilon}\right)^{2L} \mathbb{P}\{(\mathbf{z}_1, X^*X\mathbf{z}_1) \leq c + 2\varepsilon C\} + \mathbb{P}\{\lambda_{\max}(X^*X) \geq C\}.
\end{aligned}
$$

Part (ii) now follows using the result of part (i) with a sufficiently large $C$, choosing $\varepsilon > 0$ sufficiently small and using that $L/N = \nu$ is small enough. $\square$

LEMMA 7.4. *Let $X$ be a $N \times N$ Hermitian random matrix as described in (1.1) and we assume condition (1.3). Then, for $K_0 > 0$ large enough,*

$$
\mathbb{P}\{\lambda_{\max}(X) \geq K_0\} \leq e^{-c_0 K_0^2 N}
$$

*with a constant $c_0$ depending only on $\delta$ in (1.3).*

PROOF. Fix $\mathbf{z} \in \mathbb{C}^N$ with $\|\mathbf{z}\| = 1$. Then, with the notation $\mathbf{X}_j = (X_{j1}, \dots, X_{jN})^*$ for $j = 1, \dots, N$,

$$
\begin{aligned}
&\mathbb{P}\{(\mathbf{z}, X^*X\mathbf{z}) \geq C\} \\
(7.20) \qquad &\leq e^{-\kappa C N}\mathbb{E}e^{\kappa N \sum_j |\mathbf{X}_j \cdot \mathbf{z}|^2} \\
&\leq e^{-\kappa C N}\mathbb{E}e^{2\kappa N \sum_j |\sum_{l \leq j} X_{jl} \cdot z_l|^2} e^{2\kappa N \sum_j |\sum_{l > j} X_{jl} \cdot z_l|^2} \\
&\leq e^{-\kappa C N}(\mathbb{E}e^{4\kappa N \sum_j |\sum_{l \leq j} X_{jl} \cdot z_l|^2})^{1/2}(\mathbb{E}e^{4\kappa N \sum_j |\sum_{l > j} X_{jl} \cdot z_l|^2})^{1/2}.
\end{aligned}
$$

Next, choosing $\kappa > 0$ sufficiently small, we can show that, similarly to (7.13),

$$
(7.21) \qquad \mathbb{E}e^{4\kappa N \sum_j |\sum_{l \leq j} X_{jl} \cdot z_l|^2} = \prod_{j=1}^N \mathbb{E}e^{4\kappa N |\sum_{l \leq j} X_{jl} \cdot z_l|^2} \leq \text{const}^N
$$

and

$$(7.22) \qquad \mathbb{E} e^{4\kappa N \sum_j |\sum_{l \geq j} X_{jl} \cdot z_l|^2} = \prod_{j=1}^{N} \mathbb{E} e^{4\kappa N |\sum_{l \leq j} X_{jl} \cdot z_l|^2} \leq \text{const}^N$$

because $\sum_{l \leq j} |z_l|^2 \leq 1$ and $\sum_{l > j} |z_l|^2 \leq 1$. Thus, from (7.20), we have, for $C$ large enough,

$$\mathbb{P}\{(\mathbf{z}, X^* X \mathbf{z}) \geq C\} \leq e^{-c_1 C N}$$

for a constant $c_1$ only depending on $\delta$. From the last equation, the lemma follows with $C = K_0^2$ by the same argument that was used at the end of the proof of part (i) of Lemma 7.3. $\quad \square$

L. ERDŐS
B. SCHLEIN
INSTITUTE OF MATHEMATICS
UNIVERSITY OF MUNICH
THERESIENSTR. 39
D-80333 MUNICH
GERMANY
E-MAIL: schlein@math.lmu.de

H.-T. YAU
DEPARTMENT OF MATHEMATICS
HARVARD UNIVERSITY
CAMBRIDGE, MASSACHUSETTS 02138
USA